\documentclass[fleqn,usenatbib,useAMS]{mnras}

\usepackage{graphicx}   
\usepackage{amsmath}    
\usepackage{amssymb}    
\usepackage{multicol}   
\usepackage{bm}         
\usepackage{pdflscape}  
\usepackage{bigints}    
\usepackage{amsmath}    
\usepackage[euler]{textgreek} 

\newcommand{\chem}[2]{$\mathrm{^{#1}#2}$}
\newcommand{\angstrom}{\textup{\AA}}
\def\Msun{M$_\odot$\,}
\def\msun{M$_\odot$}
\defcitealias{pumo23}{Paper I}

\usepackage{mathptmx}

\title[Long-rising SNe: a new analytical model]{Long-rising Type II supernovae resembling supernova 1987A - II. A new analytical model to describe these events}

\author[M.L. Pumo \& S.P. Cosentino]{M.L. Pumo$^{1,2,3,4}$\thanks{Contact e-mail: \href{mailto: marialetizia.pumo@unict.it}{marialetizia.pumo@unict.it}} \& S.P. Cosentino$^{1,2}$\thanks{Contact e-mail: \href{mailto:stefano.cosentino@dfa.unict.it}{stefano.cosentino@dfa.unict.it}}%
\\
$^{1}$ Universit\`a degli studi di Catania, Dip.~di Fisica e Astronomia ``Ettore Majorana'', 95123, Catania, Italy\\
$^2$ INAF - Osservatorio Astrofisico di Catania, 95123, Catania, Italy\\
$^3$ Laboratori Nazionali del Sud-INFN, 95123, Catania, Italy\\
$^4$ CSFNSM c/o Dip.~di Fisica e Astronomia ``Ettore Majorana'', 95123, Catania, Italy}

\date{Accepted 2025 February 8. Received 2025 January 19; in original form 2024 August 23}

\pubyear{2025}

\begin{document}
\label{firstpage}
\pagerange{\pageref{firstpage}--\pageref{lastpage}}
\maketitle

\begin{abstract}
With the aim of improving our knowledge on supernova (SN) 1987A-like objects and, more in general, on H-rich SNe, we have developed a new analytic model to describe their post-explosive evolution. The distinctive features of this model are the possibility to evaluate the emitted luminosity and the SN expansion velocity, taking into account the recombination of the ejected material, the heating effects due to the \chem{56}{Ni} decay in the computation of the recombination front position, and the presence of an outer thin shell not-homologously expanding. In this paper, we present the model and a comparison with observations of SN 1987A, showing that its bolometric light curve and expansion velocity are accurately reproduced by the model. We also investigate the modeling degeneration problem in H-rich SNe and the possibility to ``standardize'' the subgroup of SN 1987A-like objects. Moreover we present new Ni-dependent relationships, based on our model, which link some features of the bolometric light curve of 1987A-like SNe (namely, the peak luminosity and its width) with the main physical properties of their progenitor at the explosion (i.e.~the ejected mass, the explosion energy, the progenitor radius at the explosion, and the amount of \chem{56}{Ni} present in the ejecta), showing that such relations are in excellent agreement with observations of real SNe. From our model, we also deduce new scaling relations which may be used for estimating the main SN progenitor's physical properties at the explosion, once only the photometric behaviour of the SN 1987A-like object is known.
\end{abstract}



\section{Introduction}
\label{intro}

It is commonly recognized that supernova (SN) 1987A-like objects, also known as 1987A-like SNe or  long-rising Type II events, form a subgroup of H-rich SNe having a somewhat peculiar bolometric light curve, where a secondary maximum (i.e.~after the first peak at exposion) is observed in the place of the so-called plateau \citep*[e.g.][hereafter Paper I]{turatto07,kleiser11,taddia12,takats16,singh19,pumo23}.\par

Although 1987A-like SNe are basically rare events \citep[$\lesssim 4\%$ of all Type II SNe in a volume-limited sample, see e.g.][]{pasto12}, they are relevant to different astrophysical issues. For example, SN 1987A, the prototype of the class, has been the nearest explosion since the telescope invention and has significantly improved our knowledge about the final evolutionary fate of massive stars \citep[see e.g.][]{arnett89}, shedding light also on the link between the SN explosions and the morphological properties of their remnants \citep[see e.g.][]{orlando15}. Moreover, the subsequent discoveries of events classified as 1987A-like objects, has also improved and have been improving our understanding of the SN progenitor's physical properties at the explosion (see e.g.~\citealt{pasto12}; \citealt{taddia16}; \citetalias{pumo23}, and references therein), even opening ``new'' issues about the real nature of the physical mechanisms in action during and after the explosion following the core-collapse (see e.g.~\citealt{terreran16}; \citealt{gutierrez20}; \citetalias{pumo23}). Furthermore, some 1987A-like events as SN Refsdal, that was the first gravitationally lensed SN with multiple images, have given us important constrains on the cosmic expansion rate \citep[see e.g.][]{kelly16,banklanov21}.\par
 
In spite of their importance in astrophysics, there are still basic questions to be answered on 1987A-like SNe, related to the lack of a sufficiently accuarate analytic modelling of their characteristic secondary peak. According to the standard interpretation, this peak is the result of the interplay between the cooling effects linked to the expansion of ejecta characterized by a diffusion timescale usually greater than the one of standard Type II plateau SNe, and the heating effects due to the decay of radioactive isotopes (primarily, the \chem{56}{Ni}) synthesized during the explosion and present in the ejected material \citep[e.g.][]{arnett96,PZ11,PZ13}. In particular, the ejecta expansion causes a temperature decrease that leads to the formation of a wave-front of cooling and recombination (WCR, hereafter), which moves inward (in mass) starting from the external boundary of the ejecta. In this way, the WCR divides the ejecta into two regions: (1) an inner part that is ionized, opaque, and where the opacity is dominated by the Thomson scattering, and (2) an outer zone that is optically thin, recombined, and where the opacity is approximately negligible \citep[e.g.][and reference therein]{popov93}. Differently from standard Type II plateau SNe, during this recombination phase, the radioactive energy released in the ejecta of 1987A-like events by the \chem{56}{Ni} decay contributes significantly to their thermal energy, holding up the WCR, extending the duration of the recombination phase, and affecting the peak luminosity \cite[see e.g.][]{KW09,uc11,PZ13}. Unfortunately, the analytic models commonly used to describe the peak of 1987A-like SNe and, more in general, their whole post-explosive evolution (e.g.~the models of \citealt{arnett80}, \citealt{arnett82} and \citealt{popov93}) neglect the recombination and/or the heating effects on the WCR due to the \chem{56}{Ni} decay. All of this prevents (see e.g. \citetalias{pumo23} for details): (1) a ``tout court'' usage (i.e.~without considering the applicability limits in terms of accuracy and precision) of scaling relations based on these models to easily infer the main SN progenitor's physical properties at the explosion (namely the ejected mass $M_{ej}$, the progenitor radius at the explosion $R_0$, and the total explosion energy $E$), (2) a deeply understanding of the physical origin of correlations involving spectrophotometric features of the SN at maximum, and (3) an accurate modelling and characterization of the observational features of 1987A-like SNe through analytic models. This last point is of primary importance in the context of on-going and future SNe surveys that potentially follow the evolution of thousands or more SNe, as the Vera C.~Rubin Observatory's Legacy Survey of Space and Time \citep[see e.g.][and references therein]{alves23}. Indeed, when compared to semi-analytical or hydrodynamical models, the analytic ones give us the possibility to quickly estimate the parameters describing the progenitor's physical properties at the explosion. In this sense, they are clearly favored for the analysis of large SNe samples \citep[see also][]{KK19} and, additionally, they could enable us to build up Bayesian modeling procedures devoted to directly study the error distribution of the above mentioned parameters taking into account the model accuracy \citep[see e.g.][]{hoeting99}.\par 

In this framework, we present here a new analytic model including both the recombination effects and the heating ones on the WCR due to the \chem{56}{Ni} decay. In addition, it is also possible to consider the effects linked to the presence of a not-homologously expanding outer thin shell (OTS, hereafter) that surrounds the homologously expanding ``main envelope'' (ME, hereafter) rapresenting the bulk of the ejecta (i.e.~$\gtrsim$ 99\% of the overall ejected material; cf.~\citealt{PZ11} and see Section \ref{Sec:seconda}, for further details). This new model will enable us to study the physical behavior of 1987A-like SNe, to link their observational and physical properties, to investigate the existence of correlations among their different observables, and also to perform model fitting of single events.\par

The plan of the paper is the following. The features of the new analytic model are described in Section \ref{Sec:seconda}, devoting Section \ref{Sec:terza} to the model validation. Section \ref{Sec:quarta} presents the linking between observational features and parameters describing the progenitor's physical properties at the explosion, investigating also the modeling degeneration problem in H-rich SNe and the possibility to standardize the subgroup of the 1987A-like objects. A summary with final comments about the possible applications of this new analytic model is reported in Section \ref{Sec:quinta}.\par

\section{Model description}\label{Sec:seconda}

Our new model is able to evaluate the emitted luminosity and the expansion (or photospheric) velocity for 1987A-like objects and, more in general, for H-rich SNe having in principle any kind of ``energetic source'' into their ejecta. The distinctive feature of the model is the computation of the WCR position inside the ejecta taking fully into account these source effects like, in particular, those due to a not negligible amount of \chem{56}{Ni} in the ejected material. Moreover, it is also possible to consider the effects linked to the presence of an OTS surrounding the ME. The sole ME is essentially enough in order to estimate the main parameters describing the progenitor's physical properties at the explosion with sufficient accuracy, but an OTS is necessary to correctly reproduce all the observables at early epochs\footnote{They are tipically the first 10-30 days after the explosion, when the ejecta is completely ionized and optically thick, the emission is due to the release of internal energy on a diffusion timescale, and the cooling induced by photon diffusion is negligible because the radiation diffusion timescale is much longer than the expansion timescale \citep[see also][]{PZ11}.} \citep[see e.g.][for further details]{zampieri03,PZ11}. The model is thus able to predict the emitted luminosity and the expansion velocity during the entire post-explosion evolution ranging from the breakout of the shock wave at the stellar surface up to the nebular stage (i.e.~when the ejecta has completely recombined and its energy budget is dominated by the radioactive decays of some iron group elements produced in the explosive nucleosynthesis).\par

We remark that the model adopts the same general hypotheses used in other well-known analytical models, like those presented in \citet{popov93} and \citet{arnett80,arnett82}, and an analogous formalism to simplify the comparison of the results. A detailed description of the equations, input physics, initial conditions and numerical methods used in our new model are reported in the Sections \ref{SSe:MEA} - \ref{SSe:PSL}.

\subsection{Main equations and assumptions}\label{SSe:MEA} 

The ME is the spherical and homologous expanding part of the ejecta, which accounts the bulk of the ejected material and mostly determines the electromagnetic emission of the SN event (cf.~Sections \ref{intro} and \ref{Sec:seconda}). The distance $r$ of a Lagrangian particle from the center of the ME and its outgoing velocity $v$ are \citep[see e.g.][]{arnett80,popov93}:
\begin{equation}
\label{Eq:comocord}
r(x,t)=x~R(t)\quad\text{and}\quad v(x,t)=x~v_{sc},
\end{equation}
where $x\in [0,1]$, $R(t)$, and $v_{sc}$ are respectively the dimensionless radial coordinate, the external radius at the generic time from the explosion $t$ (or, more precisely, from the so-called shock breakout), and the scale velocity of the homologous speed field. Let $v_{sc}$ be constant, $R(t)$ has to linearly evolve from the initial radius $R_0$. Indeed, from equations (\ref{Eq:comocord}) one obtains
\begin{equation}
\label{Eq:scalingf}
\frac{dR}{dt}=v_{sc}\quad\longrightarrow\quad R(t)=R_0+v_{sc}\>t\simeq v_{sc}\>t,
\end{equation}
in which $R_0$ is generally neglected for $t$ much greater than the expansion timescale $t_e\>(\equiv R_0/v_{sc}$; see also \citetalias{pumo23} and references therein$)$.\par

Assuming that the initial density of the ME is uniform and due to the homologous expansion, the mass density has to decrease uniformly over time according the following relation:
\begin{equation}
\label{Eq:density}
\rho(x,t)=\rho_0\>\frac{R_0^3}{R^3(t)}\simeq \rho_0\>\left(\frac{t_e}{t}\right)^3,
\end{equation}
where $\rho_0= 3M_{ME}/4\pi R_0^3$ is the initial density of the ME and $M_{ME}$ is the total mass of the ME\footnote{It is equal to the total ejected mass $M_{ej}$ when the OTS is not considered, and equal to $M_{ej}-M_{OTS}\lesssim M_{ej}$ where $M_{OTS}$ is the total mass of the OTS when considering the latter (see Section \ref{SSe:PSL} for further details).}. Taking into account equation (\ref{Eq:density}), the mass coordinate $m$ for a shell of radius $r$ is linked to $x$ by the following equation:
\begin{equation}
\label{Eq:mass}
m[r(x,t)]=4\pi\int_0^{r(x,t)}\rho(r')\>r'^2dr'=M_{ME}\>x^3.
\end{equation}

Also the optical opacity inside the ME $k_t$ can be described by a function of $x$ and $t$. In particular, to take into account the steep opacity variation during the SN recombination phase, the ``two-zone'' model of \citet{popov93} has been adopted. According to this model, the ME is divided into two zones (see also \citetalias{pumo23}): an outer transparent region which is recombined, characterized by a negligible opacity (i.e.~$k_t\simeq 0$) and with a temperature less than the typical hydrogen recombination temperature's value \citep[i.e.~$T< T_{ion} \simeq 5045\,$K; see][]{IN89}, and an inner opaque zone which is hotter, ionized and with opacity roughly equal to the Rosseland mean opacity $k \simeq 0.34\,$cm$^2$g$^{-1}$. The ``jump'' sphere between these two regions coincides with the WCR \citep{IP92}, whose position is time-dependent and identified by the dimensionless coordinate $x_i(t)$. As a consequence, the opacity function can be written as:
\begin{equation}
\label{Eq:opacity}
k_t(x,t)=
\begin{cases}
k\qquad x\le x_i(t),\\
0\qquad x>x_i(t).
\end{cases}
\end{equation}

Because of the deeply different physical conditions between the opaque zone and the transparent one, the outgoing luminosity of the ME $L_{ME}$ can been split into the following two terms:
\begin{equation}
\label{Eq:ME_luminosity}
L_{ME}(t)=L_{op}(t)+L_{tr}(t),
\end{equation}
where $L_{op}$ is the outgoing luminosity from the opaque region and $L_{tr}$ is the contribution due to the transparent region. However for both regions, the bolometric luminosity outgoing from a generic shell $L$ can be evaluated using the first law of thermodynamics, given by the equation
\begin{equation}
\label{Eq:firstlaw}
\frac{\partial L}{\partial m}=\Bar{\epsilon}(m,t)-P\frac{\partial}{\partial t}\left(\frac{1}{\rho}\right)-\frac{\partial \bar{e}}{\partial t},
\end{equation}
where $\bar{e}$ is the thermal energy per gram, $P$ is the pressure, and $\bar{\epsilon}$ is the rate of heating per units of gram and second. In this way, $L_{ME}$ is the integral of the equation (\ref{Eq:firstlaw}) over the entire ME.\par
 
Inside the opaque region (i.e.~for $x\le x_i$), under the radiation dominated approximation and radiative transport condition, the thermodynamic state is uniquely determined by the temperature. Indeed, the following relations are valid:
\begin{equation}
\label{Eq:radtransport}
\bar{e}=aT^4/\rho,\qquad P=aT^4/3,\qquad L=-\frac{4\pi r^2c}{3k\rho}\frac{\partial(aT^4)}{\partial r},
\end{equation}
where $c$ is the light speed and $a$ is the radiation density constant. As for the temperature, it is possible to use the relation
\begin{equation}
\label{Eq:tempprof}
T(x,t)^4=\psi_t(x)\phi(t)\>T_0^4R_0^4/R(t)^4
\end{equation}
\citep[see][for details]{arnett96}, where $T_0$ is the initial central temperature of the ME, and $\psi_t(x)$ and $\phi(t)$ are two functions denoting the spatial and temporal dependencies, respectively. The subscript ``$t$'' in the $\psi_t(x)$ function is used to indicate that, during the recombination phase, $\psi_t$ also depends on time because the WCR moves inward following the $x_i$ temporal evolution. As in \citet{popov93}, which considers a strictly adiabatic expansion with the ``radiative-zero'' boundary condition in correspondence of $x_i$, $\psi_t(x)$ can be described by the following adapted spherical Bessel function:
\begin{equation}
\label{Eq:psi_t}
\psi_t(x)=
\begin{cases}
\frac{\sin{[\pi\>x/x_i(t)]}}{[\pi\>x/x_i(t)]}\qquad\,& x\le x_i(t)\\
0\qquad\,&x>x_i(t),
\end{cases}
\end{equation}
that, throughout the diffusive stage [i.e.~when the temperature is too high for the formation of a recombination front and, consequently, $x_i(t<t_i)=1$], becomes time-independent \citep[like to constant opacity models; see e.g.][]{arnett80} and equal to $[\sin(\pi\>x)]/(\pi\>x)$. Once defined the temperature through the relations (\ref{Eq:tempprof}) and (\ref{Eq:psi_t}), considering the purely temporal term of the temperature evolution profile $\phi(t)$, the outgoing luminosity from the WCR, derived by the last of the equations (\ref{Eq:radtransport}), can be rewritten as [see Appendix \ref{Appendix A} for the complete demonstration and, in particular, equations (\ref{Eq:Eth_ini})-(\ref{Eq:Lop_dim_App})]
\begin{equation}
\label{Eq:Lop}
L_{op}(t)\equiv L(x_i,t)=\frac{E_{th}^0}{t_d}~\phi(t)~x_i(t),
\end{equation}
where
\begin{equation}
\label{Eq:Eth_td}
E_{th}^0\equiv E_{th}(0)=\int_{0}^{M_{ME}} \bar{e}(m,0)\,dm\quad\text{and}\quad t_d\equiv \frac{9kM_{ME}}{4\pi^3cR_0}
\end{equation}
are, respectively, the total initial thermal energy and the diffusion timescale (cf.~\citetalias{pumo23}). The cooling rate of the ME, which is described by the $\phi$ function, is linked to the physical conditions of the ejected material, that significantly changes when passing from the diffusive phase to the recombination one \citep[see e.g.][and reference therein]{PZ11}. In particular, the $\phi(t)$ function transforms when $L_{op}$ becomes equal to the luminosity of a Black-Body (BB, hereafter) with temperature $2^{1/4}T_{ion}$ \citep[see][]{IP92}. This marks the beginning of recombination phase $t_i$, which corresponds also to the last epoch for which the ME is totally ionized and $x_i$ is equal to 1 \citep[see also][]{popov93}. As a consequence, the relation
\begin{equation}
\label{Eq:ticond1}
L_{op}(t_i)=\frac{E_{th}^0}{t_d}~\phi(t_i)\equiv 2\pi a c \>R^2(t_i)\>T_{ion}^4
\end{equation}
is valid and the $\phi(t)$ function has to change according to the next two considerations:
\begin{itemize}
\item during the diffusive phase (i.e.~for $t<t_i$), the ME presents a uniform opacity as in the model of \citet{arnett80} and $x_i(t)$ is kept equal to 1 because the WCR position has to coincide with the ME outer boundary, so
\begin{equation}
\label{Eq:boundcond1}
t<t_i\quad\longrightarrow\quad x_i(t)=1\quad\text{and}\quad \dot{x}_i(t)\equiv\frac{dx_i}{dt}=0;
\end{equation}
\item throughout the recombination phase (i.e.~for $t\ge t_i$), the WCR moves inwards and, similarly to \cite{popov93}, $L_{op}(t)$ evolves as a BB with radius $R_i(t)=x_i(t)\> R(t)$ and constant temperature $2^{1/4}T_{ion}$ \citep[see also][]{zampieri17}, so
\begin{equation} 
\label{Eq:boundcond2}
t\ge t_i\quad\longrightarrow\quad L_{op}(t)=2\pi a c R_{i}^2(t) T_{ion}^4.
\end{equation}
\end{itemize}
Considering both these conditions (\ref{Eq:boundcond1}) and (\ref{Eq:boundcond2}), the behaviour of $\phi$ can be derived [see Appendix \ref{Appendix A} for the complete demonstration and, in particular, equations (\ref{Eq:diffresults}) and (\ref{Eq:reccond})] and, as a consequence, using relation (\ref{Eq:Lop}) the complete time evolution of $L_{op}$ can be written as
\begin{equation}
\label{Eq:Lop_complete}
L_{op}(t)=\begin{cases}
\,\left[ \frac{E_{th}^0}{t_d}+\bigintsss_{\>0}^t S(t')\,d\left(e^{t'^2/t_a^2}\right)\right]\>e^{-t^2/t_a^2}\,& t< t_i\\
\,2\pi a c v_{sc}^2T_{ion}^4\>x_i^2(t)\>t^2\,& t\ge t_i
\end{cases},
\end{equation}
in which $t_a=\sqrt{2t_et_d}$ is the characteristic luminosity  timescale for models with constant and uniform opacity (see also \citetalias{pumo23}), and
\begin{equation}
\label{Eq:source}
S(t)=\int_0^{M_{ME}}\bar{\epsilon}dm= M_{ME}\>\int_0^1\bar{\epsilon}(x,t)dx^3
\end{equation}
is the total heating energy rate at time $t$. As in \citet{popov93}, $t_i$ is set to keep $\phi$ continuous in time [see Appendix \ref{Appendix A} for further details and, in particular, cf.~equation (\ref{Eq:boundary_conditions})], but the beginning of recombination in our model is delayed because the energy contribution provided by $S(t)$ in equation (\ref{Eq:Lop_complete}) increases the ME internal energy\footnote{Moreover note that the beginning of the recombination can be estimated through the relation $t_i\simeq t_a/\sqrt{\Lambda'}$ with $\Lambda'\equiv L_a/L_{d}'$, where $L_a=2\pi a c v_{sc}^2t_a^2T_{ion}^4$ and $L_{d}'= (E_{th}^0/t_d)+\int_{0}^{s_i} S(t_a\sqrt{s'})\> e^{s'}ds'$. A similar relation is also valid for the model of \citet[][]{popov93} and was originally derived by him [see relation (25) of \citet[][]{popov93}]. However, since no extra heating mechanisms are considered in this model, the second term of $L_{d}'$ is equal to zero and, consequently, $\Lambda'$ is larger and the beginning of recombination comes before compared to what happens in our model with equal modelling parameters (see also Fig. \ref{Fig:xiprofiles}). On the other hand, $t_i\propto R_0^{1/2}\times \sqrt{1+s(S,E,R_0,M_{ME})}$ remains proportional to $R_0^{1/2}$, as in equation (25) of \citet[][]{popov93}, since the $s$ function (i.e.~the ratio between the second and the first term of $L_d'$) contains only the second order dependencies linked to the source mechanism. \label{note_ti}}, according to a mechanism analogous to the radioactive diffusive model of \citet{arnett82}. However, differently from this latter model, the presence of a WCR is considered and, in addition to what done in \citet{popov93}, the WCR evolution is derived considering the effects of extra heating mechanisms inside the ME as the presence of radioactive nuclei like the \chem{56}{Ni}. In particular, the $x_i$ time evolution has to satisfy the following differential equation [see Appendix \ref{Appendix A} for further details and, in particular, equation (\ref{Eq:recresults})]:
\begin{equation}
\label{Eq:generalxi}
x_i^2~\frac{d}{dt}\left[t\>\left(x_i^2+\frac{t^2}{3t_a^2}\right)\right]=\frac{S_i(t)}{L_a}\quad \text{with}\quad x_i(t_i)=1,
\end{equation}
where $L_a$ is equal to $2\pi a c v_{sc}^2t_a^2T_{ion}^4$ and $S_i(t)$, described by the relation
\begin{equation}
\label{Eq:sourcei}
S_i(t)=\int_{0}^{M_{i}\equiv m[r(x_i,t)]} \bar{\epsilon} dm,
\end{equation}
is the heating energy rate inside the whole opaque region (see Section \ref{SSe:sourcef} for more details). Note that both the WCR evolution and the brightness emerging from the opaque region can be computed only when the source functions ($S$ and $S_i$) are known.\par
 
In the transparent region (i.e.~for $x>x_i$), assuming a quick and local thermalization for the radiation emitted by the heating sources, the outgoing luminosity is only due to the heating term \citep[see e.g.][]{zampieri03}. In fact, equation (\ref{Eq:firstlaw}) with $e=P\simeq0$ conditions turns into
\begin{equation}
\label{Eq:lumtra}
L_{tr}(t)=\int_{M_{i}}^{M_{ME}} \bar{\epsilon} dm=S(t)-S_i(t).
\end{equation}

According to all the hypotheses described up to here, besides the dependence on the source terms, $L_{ME}$ depends on six free parameters: $R_0$, $v_{sc}$, $M_{ME}$, $k$, $T_{ion}$, and $E_{th}^0$, where $v_{sc}$ can be substituted with the kinetic energy of the whole ME $E_k$, since the relation
\begin{equation}
\label{Eq:kinetic}
E_k\simeq\frac{1}{2}\int_0^{M_{ME}}v^2(m)dm=\frac{3}{10}M_{ME}v_{sc}^2 \rightarrow v_{sc}\simeq\sqrt{\frac{10E_k}{3M_{ME}}}
\end{equation}
is valid. Considering the typical values of $k$ and $T_{ion}$ appropriate to type II SNe (i.e.~H-rich SNe like type II plateau SNe and SN 1987A-like objects) and adopting similar values of $k$ and $T_{ion}$ for all SNe of this type, the free parameters reduce to four (namely, $R_0$, $M_{ME}$, $E_k$, and $E_{th}$). Moreover, taken into consideration that the initial thermal energy is generally approximated to half of the total explosion energy $E$ which, in turn, is entirely converted into kinetic energy after the expansion time \citep[i.e.~$E_k\simeq 2E_{th}^0\simeq E$; see e.g.][]{arnett80}, the independent modeling parameters affecting $L_{ME}$ further reduce to three (namely, $R_0$, $M_{ME}$ and $E$). However, as described in Section \ref{SSe:sourcef}, other parameters can be necessary to correctly describe possible extra energy sources.

\subsection{Radioactive decay of $^{56}$Ni and source functions}\label{SSe:sourcef}

Among the possible extra energy sources, the released energy by the radioactive decay of $^{56}$Ni is the main heating mechanism during the post-explosive phases of SNe similar to SN 1987A \citep[see e.g.][]{PZ11}. In particular, the nuclear decay chain \chem{56}{Ni}$\rightarrow$ \chem{56}{Co}$\rightarrow$\chem{56}{Fe} releases an amount of energy per gram and second given by the following relation:
\begin{align}
\label{Eq:epsilon}
\epsilon(t)=\,&\epsilon_{^{56}Ni}\> \exp\left(-\frac{t}{\tau_{^{56}Ni}}\right)+ \nonumber\\
 \,&+\frac{\epsilon_{^{56}Co}\>\tau_{^{56}Co}}{\tau_{^{56}Co}-\tau_{^{56}Ni}} \> \left[\exp\left(-\frac{t}{\tau_{^{56}Co}}\right)-\exp\left(-\frac{t}{\tau_{^{56}Ni}}\right)\right],
\end{align}
where $\epsilon_{^{56}Ni}$ (set to $3.88 \times 10^{10}$ erg~g$^{-1}$~s$^{-1}$) and $\epsilon_{^{56}Co}$ (set to $7.03 \times 10^{9}$ erg~g$^{-1}$~s$^{-1}$) are, respectively, the specific energy rate released by the decay of the $^{56}$Ni and $^{56}$Co nuclei, and  $\tau_{^{56}Ni}$ (set to $7.6 \times 10^5$ s$\simeq 8.8$ d) and $\tau_{^{56}Co}$ (set to $9.6 \times 10^6$ s$\simeq 111$ d) are their decay times. Moreover, assuming that the energy mass density $\epsilon(t)$ is spatial-independent, $\bar{\epsilon}$ can be rewritten as
\begin{equation}
\label{Eq:epsbar}
\bar{\epsilon}(x,t)=\xi(x)\>\epsilon(t),
\end{equation}
where $\xi(x)$ is a function containing the spatial information about the $^{56}$Ni mass fraction distribution inside the ME.\par 

Under the above mentioned hypothesis of quick and local thermalization for the radiation emitted by the heating sources (cf.~Section \ref{SSe:MEA}), the effects of non-local trapping and leaking of the gamma radiation due to $^{56}$Ni decay are not considered \footnote{For Type II SNe, these effects are usually negligible, at least until the end of the recombination phase or beyond, being able to become sufficiently appreciable only at very later times when the ejecta density is sufficiently low that the optical depth to gamma rays throughout the ejecta is no longer $\gg 1$ \citep[see e.g.][]{zampieri17}. This is also confirmed by observational evidences linked to the gamma emission from SN 1987A, for which the gamma flux from the decay lines of \chem{56}{Co} is observed to rise distinctly above the background only about 4-5 months after the explosion, during its radioactive tail phase \citep[][]{matz88,LS90}. On the contrary, non-local gamma-ray trapping and leakage may be significant for other classes of events, such as Type IIb or Ib/c SNe \citep[e.g.][]{NOVW14}.\label{note:Ni}}. So the total heating energy rate $S$ becomes
\begin{equation}
\label{Eq:S_total}
S(t)=\int_0^{M_{ME}}\bar{\epsilon}(m,t)\,dm=M_{Ni}\>\epsilon(t),
\end{equation}
where $M_{Ni}$ is the total mass of $^{56}$Ni initially present inside the ME that, in all respects, represents another independent modeling parameter affecting $L_{ME}$ in addition to $R_0$, $M_{ME}$ and $E$ (cf.~Section \ref{SSe:MEA}).\par

As for the other source term $S_i$, it depends on the spatial distribution of the radioactive elements and the angular emission distribution of the thermalized radiation.\par
\begin{figure}
\includegraphics[angle=0,width=93mm]{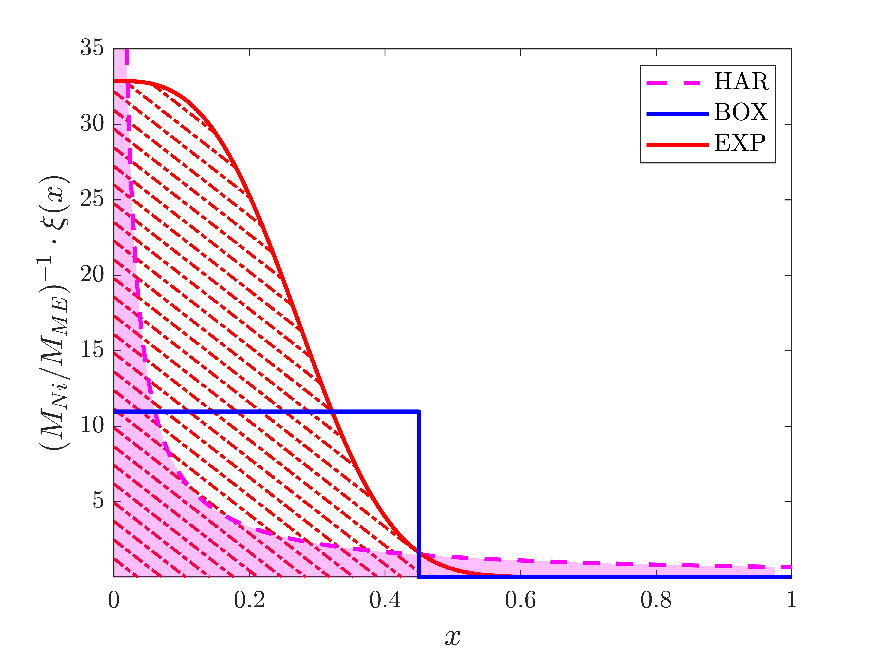}
\caption{Behaviour of the function $(M_{Ni}/M_{ej})^{-1}\>\xi(x)$ in the entire domain $x\in [0,1]$. Shaded regions under the HAR and EXP profiles contain $95\%$ of whole Ni mass.
\label{Fig:Niprofiles}}
\end{figure}

In particular, the following three spatial distributions are considered for the $^{56}$Ni mass fraction:
\begin{itemize}
\item the ``Ni Harmonic distribution'' (HAR, hereafter) given by the relation
\begin{equation}
\label{Eq:HAR}
\xi(x)=\frac{2M_{Ni}}{3M_{ME}}\>x^{-1}.
\end{equation}
In this configuration, $M_{Ni}$ is distributed according to an harmonic profile and, consequently, is weekly confined\footnote{For H-rich ejecta with very poorly confined \chem{56}{Ni}, the effects of non-local trapping and leaking of the gamma radiation could be not negligible prior of reaching the radioactive tail phase (cf.~note \ref{note:Ni}), because the assumption of complete thermalization of the radioactive energy may not hold in the outermost regions of the ejecta (i.e.~those located within about 5-10\% of the outer radius of the ejecta). However, considering that in the ``HAR'' case about 10-20\% of $M_{Ni}$ is in these outer layers, the fraction of non-thermalized gamma-rays does not exceed 2.5-5\% up to the light curve peak and 10-20\% at later phases around the end of the recombination, making the effects of non-local trapping and leaking of the gamma radiation essentially not appreciable (see also Section \ref{SubSec:quarta_2} for further details).} (see also Fig.~\ref{Fig:Niprofiles});

\item the ``Ni Box distribution'' (BOX, hereafter) for which the relation
\begin{equation}
\label{Eq:BOX}
\xi_{x_c}(x)=  \frac{M_{Ni}}{M_{ej}\>x_c^3} \times
\begin{cases}
 0 &\text{for}\quad x>x_c\\
 1 &\text{for}\quad x\le x_c
\end{cases}
\end{equation}
is valid. In this configuration, $M_{Ni}$ is uniformly confined within a sphere of radius $R_c(t)=x_c\>R(t)$, with $x_c$ being a coefficient representative of the so-called \chem{56}{Ni} mixing \citep[e.g.][]{young04}. This coefficient is set to 0.45 in order to have a \chem{56}{Ni} spatial distribution as similar as possible to the one adopted in other models reported in the literature like, in particular, those calculated with the code described in \citet*[][]{pumo10} and \citet[][]{PZ11}, that have been used to further validate and test the new model presented here;
\item the ``Ni Exponential distribution'' (EXP, hereafter) defined by the relation
\begin{equation}
\label{Eq:Ni_exp_dis}
\xi_{x_c}(x)=\frac{M_{Ni}}{M_{ej}}\>k_{dens}\> \exp{\left[-k_{mix}\left(\frac{x}{x_c}\right)^3 \right]},
\end{equation}
where $k_{mix}$ and $k_{dens}$ can be fixed using normalization conditions. Indeed, considering a sphere of dimensionless radius $x_c$ into which is confined $95\%$ of $M_{Ni}$, the distribution coefficients have to satisfy the equations
\begin{equation}
k_{mix}:\,\,\, 5\%\> e^{k_{mix}}=1-95\%\> \left[e^{(1-1/x_c^3)}\right]^{k_{mix}}
\end{equation}
and
\begin{equation}
\label{Eq:distrcoef}
k_{dens}=\,\frac{k'_{mix}}{1-\exp{\left(-k'_{mix}\right)}} \qquad\text{with}\qquad k'_{mix}=\frac{k_{mix}}{x_c^3}.
\end{equation}
\end{itemize}

Concerning the direction of the thermalized radiation, the following two assumptions are used:
\begin{itemize}
\item the ``Straight Outflow Emission'' (SOE, hereafter), according to which the thermalized radiation produced inside the transparent region contributes only to the SN brightness, while the radiation produced in the opaque region maintains hot the WCR front. So it is valid the relation
\begin{equation}
\label{Eq:SOE}
S_i(t)=M_{ME}\>\epsilon(t)\>\int_0^{x_i(t)}\xi(x)\>dx^3;
\end{equation}
\item the ``Isotropic Emission'' (IE, hereafter), according to which the thermalized radiation produced in the transparent region, being isotropically scattered, not only contributes to the SN brightness but also reaches the opaque region, contributing to sustain the WCR front. The fraction of energy reaching the opaque region depends on a geometric factor $g(x)$ --- equal to $(0.5)\left[1-\sqrt{1-(x_i/x)^2}\right]$ --- which is defined as the ratio between the solid angle of the so-called emission cone with vertex in $x$ intercepting the sphere of radius $x_i$ and $4\pi$. So, in this case, it is valid the relation
\begin{equation}
 \label{Eq:IE}
S_i(t)=M_{ME}\>\epsilon(t)\>\left[\int_0^{x_i(t)}\xi(x)\>dx^3+\int_{x_i(t)}^{1}\xi(x)g(x)dx^3\right].
\end{equation}
\end{itemize}

\begin{figure}
\includegraphics[angle=0,width=92mm]{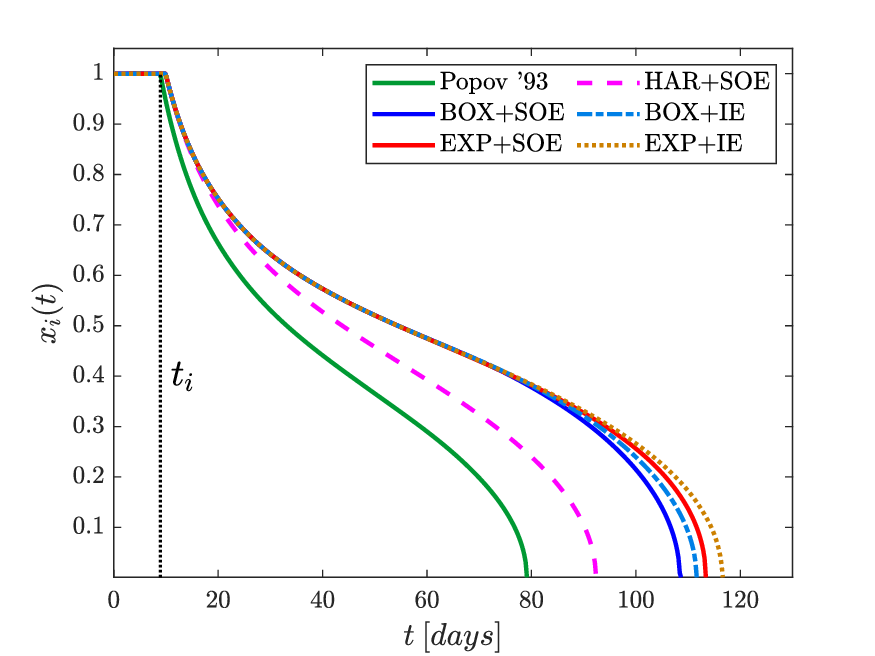} 
\caption{WCR position as a function of $t$ for different ``sub-models'' (i.e.~considering different assuptions about the \chem{56}{Ni} mass fraction distribution inside the ME and the spatial emission of the thermalized radiation; see text for further details). The modelling parameters ($E=1.3$ foe, $M_{ME}=16\,$M$_\odot$, $R_0=3$x$10^{12}\,$cm, and $M_{Ni}=0.075\,$M$_\odot$) are the same for all the sub-models, and are set equal to those inferred for SN 1987A by \citet{orlando15} through the hydrodynamical modelling that uses the code described in \citet[][]{pumo10} and \citet[][]{PZ11} for analysing the first 250 post-explosive days of SN 1987A. The WCR position of the model described in \citet{popov93} is also reported for sake of comparison. The vertical $t_i$-line marks the beginning of recombination for this model. Note that the $t_i$ value for all other Ni-dependent sub-models is about two days longer (cf.~Section \ref{SSe:MEA}).
\label{Fig:xiprofiles}}
\end{figure}

Combining all former assumptions concerning the \chem{56}{Ni} mass fraction distribution inside the ME and the spatial emission of the thermalized radiation, six different types of source function can be substituted in equation (\ref{Eq:generalxi}) (see Appendix \ref{Appendix B} for further details), resulting in as many ``sub-models'' with a different WCR time evolution (see Fig.~\ref{Fig:xiprofiles}). Throughout the manuscript, each of these sub-models is identified by two labels: the first one --- equal to ``HAR'', ``BOX'', or ``EXP'' --- refers to the adopted \chem{56}{Ni} mass fraction distribution inside the ME, while the second one --- equal to ``+SOE'' or ``+IE'' --- indicates the assumption employed to describe the spatial emission of the thermalized radiation. An additional label --- equal to ``+OTS'' --- is used to mark sub-models where the presence of an OTS is also considered, as described in Section \ref{SSe:PSL}.

\subsection{OTS contribution and expansion velocity}\label{SSe:PSL}

\begin{figure}
\includegraphics[angle=0,width=92mm]{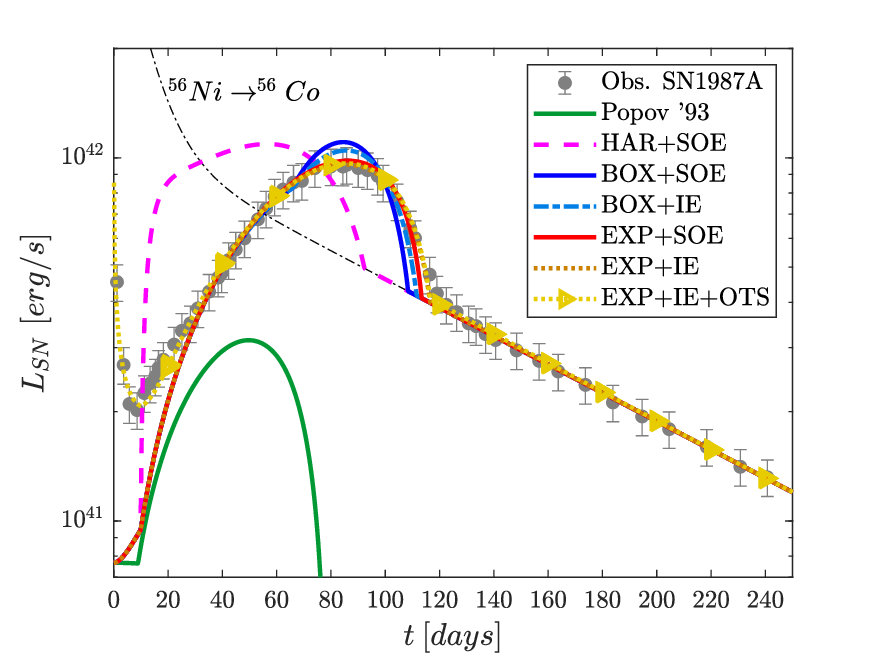} 
\caption{Bolometric luminosity of SN 1987A \citep[data taken from][]{catchpole87} and computed LCs for different sub-models considered in this work (cf.~Section \ref{SSe:sourcef}). The modelling parameters are the same described in the caption of Fig.~\ref{Fig:xiprofiles}. For the sub-model ``EXP+IE+OTS'', the parameter $\delta_0$ is put to $2.2$x$10^{-3}$ (see Section \ref{Sec:terza} for further details) and, accordingly, $M_{OTS}$ is equal to $2.7$x$10^{-2}$M$_\odot$ (see text for further details). As in Fig.~\ref{Fig:xiprofiles}, the computed LC for the model described in \citet{popov93} is also reported for sake of comparison. The luminosity due to only the \chem{56}{Ni} decay for $M_{Ni}=0.075\,$M$_\odot$ is displayed as well.
\label{Fig:bolo}}
\end{figure}

As previously mentioned (cf.~Section \ref{Sec:seconda}), an OTS surrounding the ME is necessary to correctly reproduce the SN observables like, in particular, the bolometric light curve (LC, hereafter) at early epochs. Indeed, although the OTS mass $M_{OTS}$ is negligible compared to $M_{ME}$ and, consequently, essentially irrelevant to the total ejected mass $M_{ej}$ ($\equiv M_{ME}+M_{OTS}\gtrsim M_{ME}$), its presence increases the SN brightness during the early post-explosive phase \citep*[see also][and references therein]{waxman07}, making its contribution in terms of luminosity not negligible during the first $\sim$ 10-30 d after the explosion (see also Fig.~\ref{Fig:bolo}).

In order to include the OTS and mimic its contribution in terms of luminosity in our model, we add a shell of matter at the outer border of the ME, following Waxman and co-workers for describing the shell properties \citep*[see e.g.~\citealt{waxman07}; \citealt{RW11}; \citealt{WK17};][and references therein]{morag23}. In particular, its initial thickness is $\Delta R_{0,OTS}= \delta_0 R_{0,ej}\equiv \delta_0 \> (R_0 + \Delta R_{0,OTS})= \left(\frac{\delta_0}{1-\delta_0}\right) R_0$ with $\delta_0 \ll 1$, and the main physical properties of the shell (like its density profile and velocity field) are described by power-law relationships. Furthermore, since these relations present some numerical coefficients essentially set arbitrarily, to fix such coefficients in the most physically consistent way as possible, we impose that initially (i.e.~for $t=0$) both the density profile and the velocity field inside the whole ejecta are continuous functions at the boundary between the ME and the OTS, that is at $R(t=0)=R_0$.\par

Adopting this approach, the initial density profile inside the OTS $\rho_0'$ can be described by the following function of the ejecta mass fraction $\delta_m$ lying above $R_0$ \citep[see also][and references therein]{WK17}:
\begin{equation}
\label{Eq:rho_0_dm}
\rho_0'(\delta_m)=\rho_0\>f_\rho^{\frac{1}{n+1}}\>\delta^{n/(n+1)}\simeq\rho_0\>\left(\frac{\delta_m}{\delta_m^0}\right)^{\frac{n}{n+1}},
\end{equation}
where $\delta=1-r/R_{0,ej}$, $f_{\rho}=\delta_0^{-n}$ [with $\delta_0=\delta(r=R_0)=1-R_0/R_{0,ej}$ and $n$ being an index depending on the type of energy transport mechanism equal to $3$ for radiative ejecta], and $\delta_m^0=\delta_m(r=R_0)$ is the whole mass fraction of the OTS which, for $\delta_0 \ll 1$, becomes
\begin{align}
\label{Eq:dm0}
\delta_m^0 &=\frac{M_{OTS}}{M_{ej}}=\frac{M_{ej}-M_{ME}}{M_{ej}}=\frac{1}{M_{ej}}\int_{R_0}^{R_{0,ej}}\rho_0'(r)4\pi r^2dr= \nonumber\\
&=\frac{4\pi}{M_{ej}}\int_{(1-\delta_0)R_{0,ej}}^{R_{0,ej}}\rho_0'(r)r^2dr\simeq \frac{3\>\delta_0}{n+1}.
\end{align}
In this way the relation $\rho_0'(\delta_m^0)=\rho_0$ is verified, and the initial density profile inside the whole ejecta becomes a continuous function at $R_0$, that is the place at which the initial density stops to be uniform (and equal to $\rho_0$ as inside the entire ME) and starts to decrease following a power-law (because the OTS begins). Similarly, also the velocity field inside the whole ejecta can be described by a function which is initially continuous at $R_0$ and that is a power-law relationship inside the OTS. In particular, after the shock-breakout, contrary to the ME, the OTS velocity field is no more homologous and the following relation can be used \citep[see also][and references therein]{morag23}:
\begin{equation}
\label{Eq:velo_shell}
v'(\delta_m)=v_{sc}\>\left(\frac{\delta_m}{\delta_m^0}\right)^{-\frac{n\beta}{n+1}},
\end{equation}
where $\beta= 1.19$ is a numerical coefficient linked to the shock propagation inside the OTS \citep[see e.g.][and references therein]{RW11}, and the other coefficients have been fixed so as to initially have continuity at $R_0$ and, as a consequence, to verify the relation $v'(\delta_m^0)=v(R_0)=v(x=1,t=0)=v_{sc}$. Given the velocity profile reported in equation (\ref{Eq:velo_shell}), the radial coordinate and the density profile inside the OTS evolve over the time, respectively, as
\begin{equation}
\label{Eq:r_OTS}
r(\delta_m,t)\simeq v'(\delta_m) t
\end{equation}
[similarly to what happens inside the ME; cf.~equation (\ref{Eq:scalingf})], and
\begin{equation}
\label{Eq:rho_OTS_WK17_adapted}
\rho'(\delta_m,t)=-\frac{M_{ej}}{4\pi r^2t}\left[\frac{dv'}{d(\delta_m)}\right]^{-1}
\end{equation}
\citep[see also][]{WK17}. Inserting relations (\ref{Eq:velo_shell}) and (\ref{Eq:r_OTS}) into equation (\ref{Eq:rho_OTS_WK17_adapted}), after some algebra one obtains the time-dependent OTS density
\begin{align}
\label{Eq:rhoshell_t}
\rho'(\delta_m,t) &\simeq \frac{\rho_0\> \delta_m^0\> (t_e/t)^3}{3(1-\delta_m^0)\> \beta \> [n/(n+1)]} \> \left(\frac{\delta_m}{\delta_m^0}\right)^{\frac{n(3\beta+1)+1}{n+1}}\simeq \nonumber\\
 &\simeq \frac{\rho_0\> \delta_m^0\> (t_e/t)^3}{3\beta [n/(n+1)]}\> \left(\frac{\delta_m}{\delta_m^0}\right)^{\frac{n(3\beta+1)+1}{n+1}}.
\end{align}

As for the thermal properties, the OTS mass elements traversed by the shock reach thermal equilibrium \citep[see][]{W76} and the post-shock temperature profile inside the OTS $T_0'$ can be written as  
\begin{equation}
\label{Eq:temp_shell_standard}
T_0'(\delta_m)=\left(\frac{18\rho_0'\>v_{sh}^2}{7a}\right)^{1/4}
\end{equation}
\citep[see e.g.][]{WK17}, where $v_{sh}$ is the velocity of the shock inside the OTS, linearly linked to the OTS velocity field by the relation 
\begin{equation}
\label{Eq:v_shock_standard}
v_{sh}=v'(\delta_m)/f_v,
\end{equation}
being $f_v$ a numerical coefficient $\simeq 1.71$ for OTS in SN ejecta \citep[in particular $f_v=\sqrt{10/3}\>(4\pi/3)^{-\beta}/A_v$, where $A_v$ is another numerical coefficient equal to $0.79$; for further details see e.g.][and references therein]{RW11}. Now, assuming that the OTS thermal evolution is adiabatic (i.e.~$T \propto \rho^{1/3}$), after some algebra one obtains the time-dependent OTS temperature
\begin{equation}
\label{Eq:temp_shell_t}
T(\delta_m,t)
=T_{ion}\> \frac{t_{re}}{t}\> \left(\frac{\delta_m}{\delta_m^0}\right)^{\frac{n(2\beta+1)+4/3}{4(n+1)}},
\end{equation}
where $t_{re}$ is the characteristic OTS recombination timescale given by the following relation:
\begin{align}
\label{Eq:t_r}
t_{re} &=t_e\left[\frac{18\>(n+1)^{4/3}\rho_0\>v_{sc}^2}{7(3\beta\> n/\delta_m^0)^{4/3}f_v^2\>aT_{ion}^4}\right]^{1/4}\propto R_{0,ej}^{1/4} M_{ej}^{1/2} E^{-1/4} \delta_0^{1/3}\simeq \nonumber\\
&\simeq R_0^{1/4} M_{ME}^{1/2} E^{-1/4} \delta_0^{1/3}.
\end{align}
The OTS recombination occurs when the temperature inside the OTS goes below $T_{ion}$. In this case, the position of the WCR inside the OTS $\delta_m^i$ is such that $T(\delta_m^i,t)=T_{ion}$ and, according to equation (\ref{Eq:temp_shell_t}), it evolves over the time as
\begin{equation}
\label{Eq:delta_i}
\delta_m^i(t)=\delta_m^0\> \left(\frac{t}{t_{re}}\right)^{\frac{4(n+1)}{n(2\beta+1)+4/3}}.
\end{equation}

Due to the low-density condition, the approximation of overlapping between WCR and the photosphere used for the ME (cf.~Section \ref{SSe:MEA}), could not be reasonable inside the OTS. For this reason, to individuate the photosphere's position in the latter case, the optical depth $\tau_s$ is first evaluated similarly to \cite{RW11} as
\begin{align}
\label{Eq:opt_depth}
\tau_s(\delta_m,t) &= \frac{M_{OTS}}{4\pi}\int_{0}^{\delta_m}k_t\frac{d\delta_m'}{r^2(\delta_m',t)}=\frac{kM_{OTS}}{4\pi}\int_{\delta_m^i}^{\delta_m}\frac{d\delta_m'}{r^2(\delta_m',t)}=\nonumber\\
&=\left(\frac{t_{ph}}{t}\right)^2\times \left[\left(\frac{\delta_m'}{\delta_m^0}\right)^{\frac{n(2\beta+1)+1}{n+1}}\right]_{\delta_m^i}^{\delta_m},
\end{align}
 where
\begin{equation}
\label{Eq:t_ph}
t_{ph}=\sqrt{(kM_{OTS}\delta_m^0/4\pi v_{sc}^2)(n+1)/[n(2\beta+1)+1]},
\end{equation}
and doing so, the photosphere is located in $\delta_m^{ph}$ such that $\tau_s(\delta_m^{ph},t)=1$, obtaining the following relation:
\begin{equation}
\delta_m^{ph}(t)=\delta_m^0\> \left[\left(\frac{t}{t_{ph}}\right)^2+\left(\frac{t}{t_{re}}\right)^{4\frac{n(2\beta+1)+1}{n(2\beta+1)+4/3}}\right]^{\frac{n+1}{(2\beta+1)n+1}}.
\end{equation}
Once known $\delta_m^{ph}$, the OTS luminosity $L_{OTS}$ can be approximated as a BB of radius $\delta_m^{ph}$ and temperature $T(\delta_m^{ph})$. This luminosity contributes to the whole SN luminosity $L_{SN}$ until the OTS recombination starts. After that, its contribution quickly decreases and the ME brightness becomes predominant (see also Fig.~\ref{Fig:bolo}). To take into account this effect, an exponential factor with $t_{re}$ as a timescale has been adopted to infer $L_{SN}$ (see Section \ref{Sec:terza} for further details), that can be written as 
\begin{equation}
\label{Eq:L_SN}
L_{SN}(t)=L_{ME}(t)\> \left(1-e^{-t/t_{re}}\right)+L_{OTS}(t)\> e^{-t/t_{re}},
\end{equation}
where
\begin{equation}
\label{Eq:L_OTS}
L_{OTS}(t)=\pi a c\> \> v'^2[\delta_m^{ph}(t)]\> t^2\> T^4[\delta_m^{ph}(t),t].
\end{equation}
  
A similar semi-empirical approach can be adopted to describe the SN expansion velocity $v_{SN}$, whose general expression can be thus written as
\begin{equation}
\label{Eq:velocity}
v_{SN}(t)=v_{ME}(t)\> \left(1-e^{-t/t_{re}}\right)+ v_{OTS}(t)\> e^{-t/t_{re}},
\end{equation}
where $v_{ME}$ and $v_{OTS}$ are the expansion velocity of the ME and the OTS, respectively. Inside the ME, to a first approximation (see below for a more in-depth treatment), since the WCR position, the photosphere's position, and the line formation region\footnote{This region refers to the layers of the ejecta responsible for the absorption component of the P-Cygni profile for the lines used to estimate the expansion velocity in observed SN events \citep[see e.g.][for details]{KK74}.} roughly coincide, $v_{ME}(t)$ can be well approximated with the WCR velocity. Consequently the relation
\begin{equation}
\label{Eq:velocity_ME}
v_{ME}(t)=v_{sc}\> x_i(t) 
\end{equation}
is valid [cf.~equation (\ref{Eq:comocord}) for $x=x_i$]. Instead, inside the low-dense OTS, the approximation of an overlapping among the above mentioned three zones (i.e.~WCR, photosphere, and line formation region) it is not valid. So, in order to well simulate the OTS expansion velocity and accurately reproduce the velocities observed in real SN events, is necessary to individuate the line formation region. Considering that, in addition to being above the photosphere (because the opacity in the continuum is smaller than the one in the line), the line formation region is generally also above the WCR position \citep[see also][]{DH05}, it is possible to adopt the following relation:
\begin{equation}
\label{Eq:velocity_OTS}
v_{OTS}(t)= v'[\eta\> \delta_m^i(t)],
\end{equation}
where $\eta$ --- put to $0.1$ in the calculations presented in this paper (see Section \ref{Sec:terza} for further details) --- is a dimensionless parameter describing the separation degree between the line formation region and the WCR inside the OTS.\par

\begin{figure}
\includegraphics[angle=0,width=91mm]{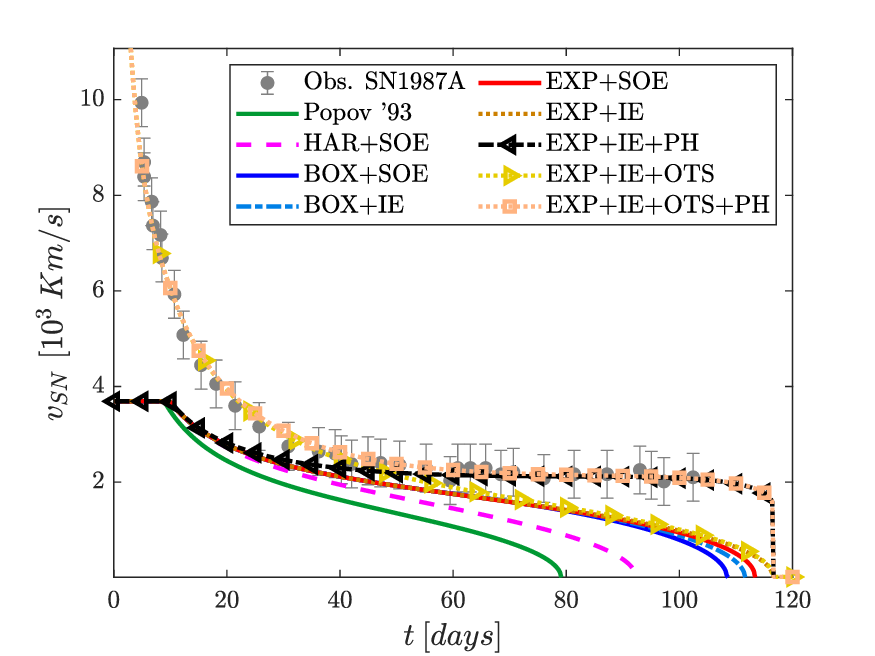} 
\caption{Expansion velocity (estimated from the FeII-5169\AA\, line) of SN 1987A \citep[data taken from][]{phillips88} and computed velocities for different sub-models considered in this work. The modelling parameters are the same described in the caption of Fig.~\ref{Fig:bolo}. Sub-models ``EXP+IE+OTS'' and ``EXP+IE+OTS+PH'' take into account the OTS effects, using equation (\ref{Eq:velocity}) to evaluate the SN expansion velocity. In the other sub-models, it is evaluated considering only the ``ME contribution'' [i.e.~according to the relation $v_{SN}\simeq v_{ME}$, with $v_{ME}$ evaluated from equation (\ref{Eq:velocity_ME_HP}) for the sub-model ``EXP+IE+PH'' and from equation (\ref{Eq:velocity_ME}) for the remaining sub-models]. The value of the line opacity for the FeII-5169\AA\, line adopted in this paper is $k_{FeII-5169\angstrom}= 4.5$x$10^{-3}\,$cm$^2$g$^{-1}$, and has been determined fitting the ``EXP+IE+OTS+PH'' sub-model on the observed expansion velocity of SN 1987A (see Section \ref{Sec:terza} for further details).
\label{Fig:velo}}
\end{figure}
\begin{figure*}
\includegraphics[angle=0,width=170mm]{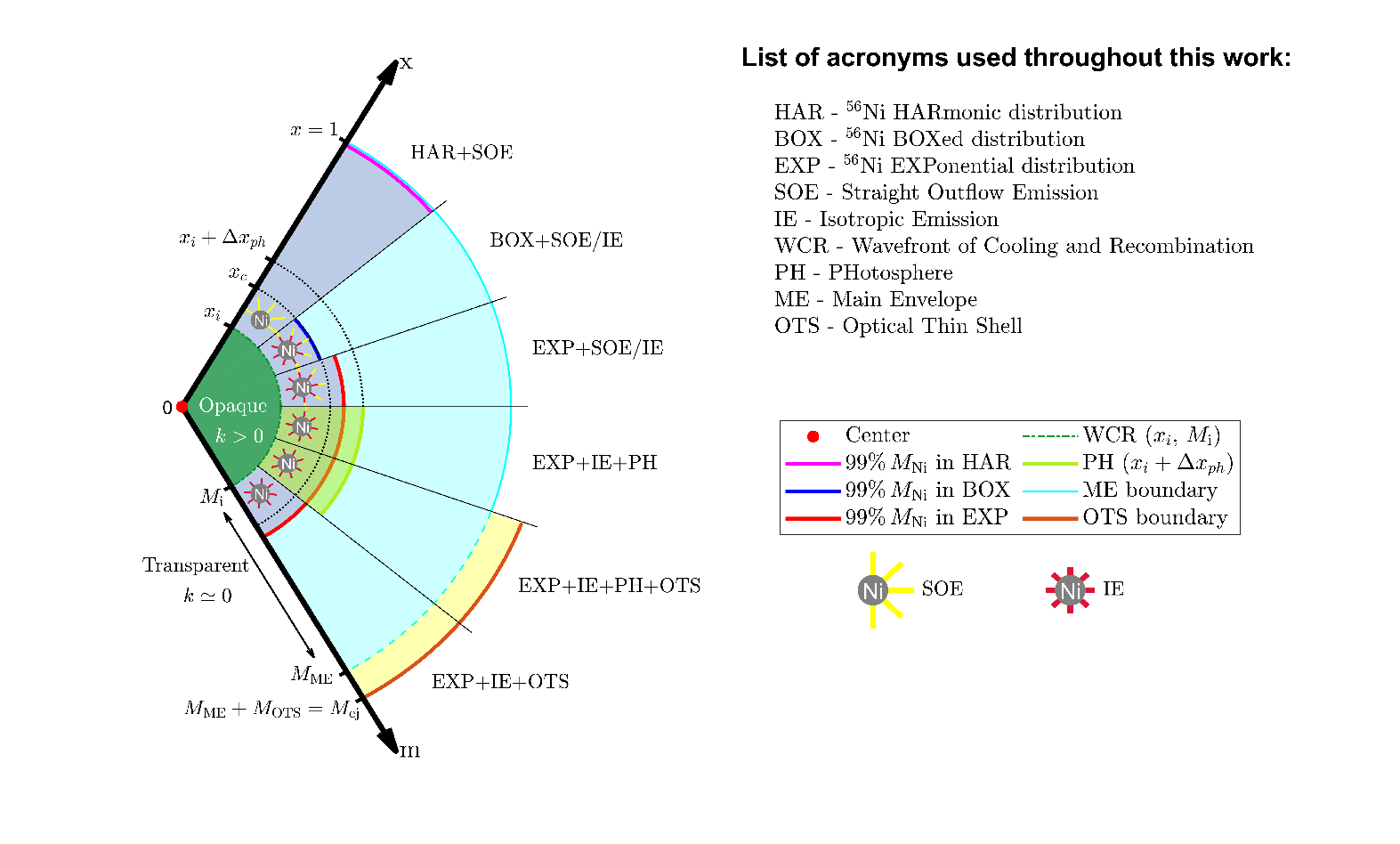} 
\caption{Schematic structure (not in scale) of the SN ejecta according to the modelling hypotheses adopted in the different sub-models, with a list of the widely used acronyms.
\label{Fig:sub_models}}
\end{figure*}

This not-negigible spatial separation between WCR and the line formation region doesn't only interest the OTS and therefore has to be taken into account when estimating the SN expansion velocity at early post-explosive phase, but it might be relevant also at later phases \citep[see e.g.][]{Inserra12,Inserra13}. Indeed, also inside the ME, under opportune physical conditions \citep[e.g.~when the density is sufficient low or due to the variations of ionization balance processes and opacity; see also][for further details]{PZ11}, the line formation region cannot be well approximated with the WCR position or the photosphere's position, being well above them. In this case, it is useful to analyse the behavior of the optical depth $\tau_{S,l}$ for the line used to estimate the expansion velocity. In particular, since the ME is an expanding structure, the relation 
 \begin{equation}
 \label{Eq:sobolev}
  \tau_{S,l}=\chi_l\> \Delta l_S= k_{l}\> \rho\> c(\Delta \nu/\nu)\> [R(t)/v_{sc}]
 \end{equation}
 is valid \citep[see][]{sobolev}, where $\tau_{S,l}$ is linked to the extinction coefficient of the line $\chi_l$ ($\equiv k_l\> \rho$, being $k_l$ the line opacity) and the Sobolev length $\Delta l_S$ [$\equiv c(\Delta\nu/\nu)\> (dv/dr)^{-1}$], whose product even depends on the line Doppler shift $\Delta\nu/\nu$. Considering the velocity field topology inside the ME, the relative motion between two shells spaced $\Delta x$ produces a relative frequency shift between them of $v_{sc}(\Delta x/c)$. Inserting this relative frequency shift in the term $\Delta\nu/\nu$ of equation (\ref{Eq:sobolev}), the condition $\tau_{S,l}=1$ make it possible to estimate the comoving distance between the WCR (or, more in general, the photosphere's position) and the line formation region $\Delta x_{ph}$, that can be written as
 \begin{equation}
 \label{Eq:correction_velocity_ME_HP}
 \Delta x_{ph}(t)= (t/t_e+1)^2/(k_{l}\> \rho_0\> R_0),
 \end{equation}
 so the expansion velocity associated to the considered line becomes:
 \begin{equation}
 \label{Eq:velocity_ME_HP}
 v_{ME}(t)=v_{sc}\> [x_i(t)+\Delta x_{ph}(t)]\qquad \text{until}\qquad  x_i>0.
 \end{equation}

Thus, the correction $\Delta x_{ph}$ should be taken into account to accurately reproduce the expansion velocity at late pheses (i.e.~from 40-50 days after the explosion onwards; see Fig.~\ref{Fig:velo}). Throughout the manuscript, sub-models considering such correction, are identified by the further label ``+PH''. Fig.~\ref{Fig:sub_models} shows in a schematic way the SN ejecta's structure according to the modelling hypotheses employed in the different sub-models, with a list of the main acronyms used in the manuscript and, in particular, of those adopted to identify the sub-models.

\section{Model validation \& comparison among the different sub-models}\label{Sec:terza}

SN 1987A is used to validate our new model described in Section \ref{Sec:seconda} and compare the various sub-models, as well as set the parameters $\delta_0$, $\eta$, and $k_{FeII-5169\angstrom}$ (cf.~Section \ref{SSe:PSL}). Note that $\delta_0$ determines $M_{OTS}$, which is the additional independent modeling parameter of our new model (together with the other four parameters $M_{ME}$, $E$, $R_0$, and $M_{Ni}$; cf.~Section \ref{Sec:seconda}) to be considered when taken into account the OTS contribution.\par

\begin{figure}
\includegraphics[angle=0,width=90mm]{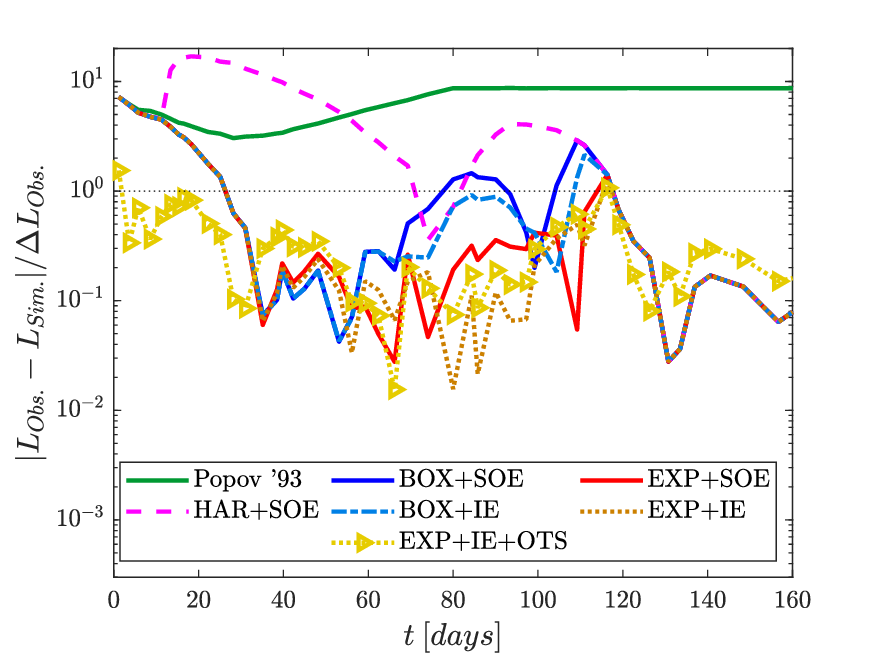} 
\caption{Absolute discrepancy between the bolometric luminosity of SN 1987A $L_{Obs.}$ and the computed one $L_{Mod.}$ as a function of time from the explosion, for the different sub-models considered in this work and reported in Fig.~\ref{Fig:bolo}. The values are normalized to the observational uncertainty of the luminosity $\Delta L_{Obs.}$. The model predictions outside the error bar of the observed bolometric luminosity are located above the horizontal dotted line. The normalized absolute discrepancy between the bolometric luminosity of SN 1987A and the computed one for the model described in \citet{popov93} is also reported for sake of comparison.
\label{Fig:bolo_chi}}
\end{figure}
\begin{figure}
\includegraphics[angle=0,width=90mm]{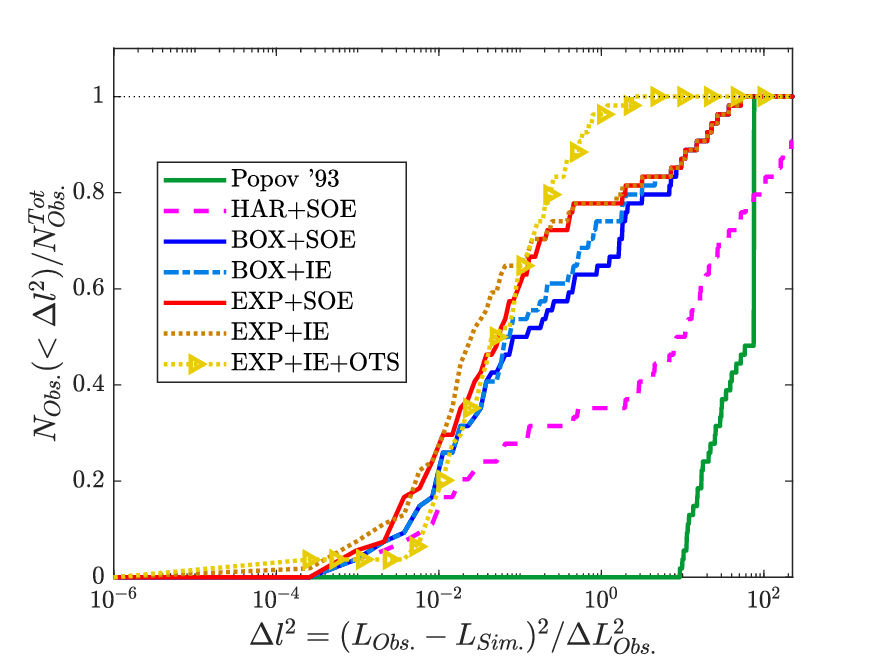} 
\caption{As for Fig.~\ref{Fig:bolo_chi}, but for the normalized cumulative frequency analysis of the squared deviation $\Delta l^2= (L_{Obs.}-L_{Mod.})^2/\Delta L_{Obs.}^2$. It is essentially equal to the normalized frequency of occurrence of squared deviations values less than a fixed $\Delta l^2$ value. The area between the curve referring to a given sub-model and the horizontal dotted line is the $\chi^2$ value of the sub-model. The normalized cumulative frequency analysis for the model described in \citet{popov93} is also reported for sake of comparison.
\label{Fig:bolo_accuracy}}
\end{figure}

As for the validation of our new model, since this work is mainly devoted to 1987A-like events, we primary point our attention on the LC. Indeed, there are no significant degeneration problems linked to the LC modelling for this subgroup of events and, consequently, the modelling of the sole LC (i.e.~without considering additional information taken from the expansion velocity modelling) is sufficient to retrieve information about the physical properties of their progenitors at the explosion (see Section \ref{SubSec:quarta_1} for details). As shown in Fig.~\ref{Fig:bolo}, the comparison with the LC of SN 1987A is very acceptable and all the sub-models are able to reproduce its main features (i.e.~ peak luminosity and phase at maximum). Moreover, the normalized discrepancy between the bolometric luminosity of SN 1987A and the computed LCs is always less than 1 during most of the post-explosive evolution (see Fig.~\ref{Fig:bolo_chi}). The agreement can be considered fully satisfactory also because we do not perform any ``fine-tuning'' of the modelling parameters $E$, $M_{ME}$, $R_0$ and $M_{Ni}$, but we adopt the values inferred by \citet{orlando15} through the hydrodynamical modelling (HM, hereafter) that uses the code described in \citet[][]{pumo10} and \citet[][]{PZ11} for analysing the first 250 post-explosive days of SN 1987A.\par 

The only exception is the HAR case with a computed LC having an earlier (and broader) peak than the one observed for SN 1987A (see curve labeled ``HAR+SOE'' in Fig.~\ref{Fig:bolo}). This is due to the harmonic profile of the \chem{56}{Ni} mass fraction distribution inside the ME adopted in this case, which leads to a less confined \chem{56}{Ni} mass than the other distributions (i.e.~BOX and EXP cases; see Fig.s \ref{Fig:Niprofiles} and \ref{Fig:sub_models}). So, the energy released by the \chem{56}{Ni} decay is quickly emitted outside the ME, explaining the initial (i.e.~around 10-50 days after the explosion) luminosity eccess compared with all other sub-models. However, the peak luminosity of SN 1987A is roughly well reproduced, contrary to what happens for analytic models that do not include source terms due to the \chem{56}{Ni} decay as, for example, that described in \citet[][]{popov93} (see curve labeled ``Popov '93'' in Fig.~\ref{Fig:bolo}). Therefore, since HAR is the easiest approach which includes \chem{56}{Ni} source terms thus also allowing to obtain a ``pure'' analytic solution for the $x_i$ time evolution (cf.~Appendix \ref{Appendix B}), it can be very useful to understand the physical origin of correlations involving spectrophotometric features at LC maximum and, more in general, to describe the physical behavior of 1987A-like objects (see Section \ref{Sec:quarta} for more details). Nonetheless, sub-models using the HAR approach are not very suitable for performing model fitting of single observed SNe events and, in this case, the usage of more accurate sub-models is more appropriate.\par

As shown in Fig.~\ref{Fig:bolo_accuracy}, the accuracy of sub-models grows up with their complexity and the sub-model ``EXP+IE+OTS'' is the most accurate one. Moreover the accuracy of the sub-models is primarily determined by the spatial \chem{56}{Ni} distribution and to a second extent by the other ``physical inputs'' as the direction of the thermalized radiation and the OTS inclusion. In particular, the direction of the thermalized radiation influences both the WCR and LC evolution, especially during the last 30-40 days before the recombination ending ($t_f$, hereafter), which occurs when $x_i\simeq 0$. In this respect, the IE approach seems to be the most realistic one (compared to the easier SOE approach), increasing $t_f$ (see Fig.~\ref{Fig:xiprofiles}) and well reproducing the LC of SN1987A around 100-120 days after the explosion (see Fig.~\ref{Fig:bolo_chi}). However, when using the IE approach, the recombination stage is extended only by about $3\%$, with the disadvantage of a greater computational cost. This is particularly noticeable when the IE approach is used in combination with the EXP approach, because the sub-model ``EXP+IE'' does not present an analytic source function (see Appendix \ref{Appendix B} for more details) and therefore a double numerical integration is needed, significantly increasing the computing time. So the sub-model ``EXP+SOE'' appears to be the best compromise between accuracy and computing time demands, at least when focusing only on estimating the physical properties of the SN progenitor at the time of the explosion without requiring an accurate modeling of the LC evolution at the early post-explosive phase. On the contrary, to accurately reproduce the early LC behavior, it is necessary to consider the OTS contribution. In this case, we stress that other ancillary hypotheses and the fifth independent modeling parameter $M_{OTS}$ (depending on $\delta_0$) have to be added (cf.~Section \ref{SSe:PSL}) to those reported in Sections \ref{SSe:MEA} and \ref{SSe:sourcef}. Unlike the other independent modeling parameters, $M_{OTS}$ can not be fixed by HM approaches and it has to be determined by a fitting procedure on the early LC. In particular, for the case of SN1987A presented in this paper, we use observations of the first 30 post-explosive days, obtaining a value of $\delta_0$ equal to $2.2$x$10^{-3}$, which implies a $M_{OTS}$ of $2.7$x$10^{-2}M_\odot$, in agreement with the post-shock breakout model of \citet{WK17}.\par

As for the expansion velocity, despite we are primary interested in the LC modelling, we point out that the sub-model ``EXP+IE+OTS+PH'' is able to accurately reproduce the observed expansion velocity of SN 1987A during the entire post-explosive evolution (see Fig.~\ref{Fig:velo}). All other sub-models are able to roughly reproduce its behavior around 30-80 days after the explosion, but they systematically  underestimate the expansion velocity at early-time (i.e.~during the first 20-30 post-explosive days), at late-time (i.e.~for $t\gtrsim 80-100$ days), or at both phases. Therefore, in order to accurately reproduce the expansion velocity, both the OTS contribution and the ``PH'' correction should be taken into account. Indeed, as expected (and already discussed in Section \ref{SSe:PSL}), the OTS contribution is responsible of the velocity boost at early-time, while the ``PH'' correction becomes fundamental  at late-time (cf., in particular, curves labeled ``EXP+IE+OTS'', ``EXP+IE+PH'', and ``EXP+IE+OTS+PH'' in Fig.~\ref{Fig:velo}).\par

Similarly to what done for the parameter $\delta_0$ when performing LC modeling considering the OTS contribution, the parameters $\eta$ and $k_{FeII-5169\angstrom}$ have also to be fixed by fitting procedures, when performing expansion velocity modelling cosidering both the OTS contribution and the ``PH'' correction. In particular, for the comparison of model predictions with observations, we use the expansion velocities of SN 1987A estimated from the FeII-5169\AA\, line. Note that $k_{FeII-5169\angstrom}$ (and, more in general, the line opacity for the line used to estimate the expansion velocity) physically depends on the strength of the line and the abundance of the relative ion above the WCR, although it can be considered approximately constant inside the ejecta as assumed in relation (\ref{Eq:correction_velocity_ME_HP}).\par

Note finally that, to our knowledge, in literature there is no analytic or numerial treatment able to accurately describe the transition of the photosphere's position from the OTS and to the ME from a physical point of view. It is probabily due to the huge variation of the involved mass scales, that prevents a unique and consistent parametrization of the entire ejecta evolution. So, in this paper, the exponential approach in relations (\ref{Eq:L_SN}) and (\ref{Eq:velocity}) is essentially adopted on semi-empirical bases, allowing us to ``join'' the ME and OTS descriptions without having discontinuities in the temporal evolution of the emitted bolometric luminosity and SN expansion velocity.

\section{Light Curves physical behavior}\label{Sec:quarta}

After analysing the dependability of our new model in Section \ref{Sec:terza}, we use this model for studying the link between the observational features of 1987A-like SNe and the main parameters describing the physical properties of their progenitors at the explosion, pointing particular attention to the LC behavior. However, prior to performing and presenting such study (see Sections \ref{SubSec:quarta_2} and \ref{SubSec:quarta_3}), in Section \ref{SubSec:quarta_1} we examine the so-called modeling degeneration problem or parameters degeneration problem \citep[see][and references therein]{pumo17,GBP19}, when modeling the observational features and, in particular, the LC features of 1987A-like events. Indeed, the possibility of reproducing essentially the same LC with more than one set of parameters, is one of the major problems to be evoided for having a robust inference of the above mentioned parameters describing the SN progenitor's physical properties and, in particular, when studying the link between such parameters and the observational features of the modelled events.

\subsection{Modeling degeneration problem and ``standardization'' of 1987A-like objects}\label{SubSec:quarta_1}

To understand when the modeling degeneration problem can arise and how it can be relevant, it is useful to analyse the mathematical dependencies between the independent modeling parameters and the free coefficients of the equations describing the post-explosive ejecta evolution like, in particular, those concerning the WCR evolution and the LC behavior. To do so, we consider the case ``EXP+SOE'' (which is the preferable sub-model in terms of best compromise between accuracy and computing time demands, when studying the link between the LC evolution and the modeling parameters; cf.~Section \ref{Sec:terza}) and perform a change of variables, introducing the following new functions:
$$ y\equiv t/t_a \quad\text{and}\quad z(y)\equiv x_i(t_a\times y).$$ 
In this way, the Cauchy problem for the WCR evolution defined by the relation (\ref{Eq:generalxi}) [see also equation (\ref{Eq:exp_diff}) for further details], can be rewritten as 
\begin{equation}
\label{diff_eq_mat}
\frac{dz}{dy}=-\frac{1}{2zy} \left[y^2+z^2-\lambda e^{-k_1y}\left(\frac{1-e^{-k_2\>z^3}}{z^2}\right)\right]\,\text{with}\, z(y_0)=1,
\end{equation}
where $k_2=k_{mix}'$ is a fixed coefficient\footnote{In principle $k_2$ should be included among the free coefficients but, in our approach, its value is fixed because it depends on the coefficient $x_c$ which, in turn, is set to 0.45 (cf.~Section \ref{SSe:sourcef}). However, note that a change in $k_2$ generally produces only secondary effects on the LC behavior.}, while $\lambda$, $k_1$, and $y_0$ are free coefficients linked to the modeling parameters and defined by the following relations:
\begin{equation}
\label{Eq:y_0}
y_0=t_i/t_a\propto \left(R_0\right)^{1/2}\times\left(E/M_{ME}^3\right)^{1/4},
\end{equation}
\begin{equation}
\label{Eq:lambda}
\lambda=q'\propto \left[M_{Ni}\times \left(M_{ME}/E\right)\right]\times\left(E/M_{ME}^3\right)^{1/2},
\end{equation}
\begin{equation}
\label{Eq:k_1}
k_1=t_a/\tau_{^{56}Co}\propto \left(E/M_{ME}^3\right)^{-1/4}.
\end{equation}
Consequently, the WCR evolution can be uniquely determined by a triplet ($y_0,\lambda,k_1$), but the modelling parameters are four (namely, $E$, $M_{ME}$, $R_0$, and $M_{Ni}$). This implies the presence of a degeneration because different 4-tuples of modelling parameters can produce the same triplet ($y_0,\lambda,k_1$) and, consequently, the same WRC evolution. \par
Although the WCR evolution is degenerate, the LC behavior is not. Indeed, inserting relations (\ref{Eq:Lop_complete}) and (\ref{Eq:lumtra}) into equation (\ref{Eq:ME_luminosity}), considering the above mentioned change of variables, and using also relations (\ref{Eq:S_total}), (\ref{Eq:EXP+SOE}) and (\ref{Eq:decay_approx}), the whole SN luminosity can be rewritten as
\begin{equation}
\label{Eq:L_SN_y}
L_{SN}(y)\simeq L_a \times \left\{y^2z^2 - \lambda e^{-(k_1y+k_2)}\times \left[1-e^{k_2(1-z^3)}\right]\right\},
\end{equation}
where $L_a= 2\pi a c v_{sc}^2t_a^2T_{ion}^4\propto M_{Ni}\times \lambda^{-1}$ [cf.~also note \ref{note_ti}] breaks the degeneration thanks to the direct dependence on $M_{Ni}$. This implies that different 4-tuples of modeling parameters produce a different LC behavior and viceversa. So the modelling of the sole LC behavior is in principle sufficient to constrain all modeling parameters.

However, for values of $\lambda$ sufficiently low (i.e. $\lambda \lesssim 10^{-2}-10^{-3})$, degeneration problems can arise when modelling the LC, because the second term on the righ-hand side of relation (\ref{Eq:L_SN_y}) becomes negligible and, consequently, the LC behavior gets essentially determined only by two free parameters. In this case, additional information of spectroscopic nature, such as that retrieved by modelling the expansion velocity, has to be also used to uniquely constrain the modelling parameters.

On the other hand, real 1987A-like events are characterized by higher values of $\lambda$, which naturally broke the LC degeneration. So no significant degeneration problems arise when performing their LC modelling. Nevertheless, the additional information of spectroscopic nature could be used to constrain all modelling parameters describing a real 1987A-like SN when its distance (or equivalently its absolute bolometric luminosity) is not known. Indeed, using the information on the expansion velocity to infer the scale velocity $v_{sc}$ [proportional to $(E/M_{ME})^{1/2}$; cf.~relation (\ref{Eq:kinetic})], it is possible to characterize the event because the following relations are valid:
$$R_0\propto y_0^2k_1^2\qquad M_{ME}\propto v_{sc}^2k_1^4\qquad E\propto v_{sc}^6k_1^8\qquad M_{Ni}\propto v_{sc}^2\lambda k_1^{-2}.$$
Moreover, since the expansion velocity measurements can be affected by the host galaxy's redshift, but they do not depends on the cosmological parameters (necessary to distance inference), a further corollary of this type of characterization may be the possibility of standardizing the 1987A-like events using spectrophotometric information. This is in agreement with what found in \citet[][]{PZ13}, according to which a purely photometric based standardization of these objects appears difficult to be realized.

\subsection{Ni-dependent relations for the peak luminosity and its width}\label{SubSec:quarta_2}

The LC peak of 1987A-like SNe, as well as the plateau for type IIP SNe, is typically described in terms of two main observational quantities: the peak (or plateau for type IIP SNe) luminosity ($L_p$, hereafter) and its width (or duration for type IIP SNe), expressed by a time interval or a characteristic epoch that is relatively easily measurable and usually linked to $t_f$ \citep[see e.g.][]{popov93,PZ13}. Through relatively simple relations derived from the analysis of ``synthetic'' LCs based on (semi-)analytic or numerical models, the values of $L_p$ and $t_f$ are also related to the parameters describing the SN progenitor at explosion (see e.g. \citealt{arnett96}; \citealt{popov93}; \citealt{PZ13}; \citealt{sukhbold16}; \citealt{KK19}; \citetalias{pumo23}).\par

Usually (in particular for $L_p$), the considered parameters are the sole $E$, $M_{ME}$, and $R$; while the parameter $M_{Ni}$ is neglected because the heating effects linked to the \chem{56}{Ni} are not considered. However, as shown in \citetalias{pumo23}, the dependence on $M_{Ni}$ must be also taken into consideration in order to have accurate relationships for 1987A-like SNe. So, using our new model which is able to take fully into account the \chem{56}{Ni} effects, it is possible to derive how $L_p$ and $t_f$ depend in detail on all the four modelling parameters that primarily determine the LC evolution (i.e.~$E$, $M_{ME}$, $R$, and $M_{Ni}$). In particular, to do so, we use our ``best'' sub-model for these purposes (i.e.~sub-model ``EXP+SOE''; cf.~Sections \ref{Sec:terza} and \ref{SubSec:quarta_1}) and make the hypothesis that both $L_p$ and $t_f$ can be written as the sum of two functions, obtaining:
\begin{equation}
\label{Eq:Lt_2comp}
\begin{cases}
t_f= t_f^0(R_0,E,M_{ME})+t_f'(M_{Ni},R_0,E,M_{ME})\\
L_p= L_p^0(R_0,E,M_{ME})+L_p'(M_{Ni},R_0,E,M_{ME}),
\end{cases}
\end{equation}
where $t_f^0$ and $L_p^0$ depend solely on $R_0$, $E$, and $M_{ME}$; while $t_f'$ and $L_p'$ depend also on $M_{Ni}$, becoming zero when $M_{Ni}= 0$.\par 

Since the model described in \citet{popov93} is a special case of our model when the $S_i(t)$ is null [see Appendix \ref{Appendix B} and, in particular, equation (\ref{Eq:Popov_eq15}) for the demonstration], $L_p^0$ and $t_f^0$ correspond to the quantities $L_{bol}(t_m)$ and $t_p$ presented in equations (26) and (27) of \citet{popov93}, respectively. Consequently they exhibit the following well-known dependencies on the modelling parameters:
 \begin{equation}
 \label{Eq:Popov_scaling}
 \begin{cases}
 t_f^0\propto R_0^{1/6}\> E^{-1/6}\> M_{ME}^{1/2}\\
 L_p^0\propto  R_0^{2/3}\> E^{5/6}\> M_{ME}^{-1/2}.
 \end{cases}
 \end{equation}
 
\begin{figure}
\includegraphics[angle=0,width=90mm]{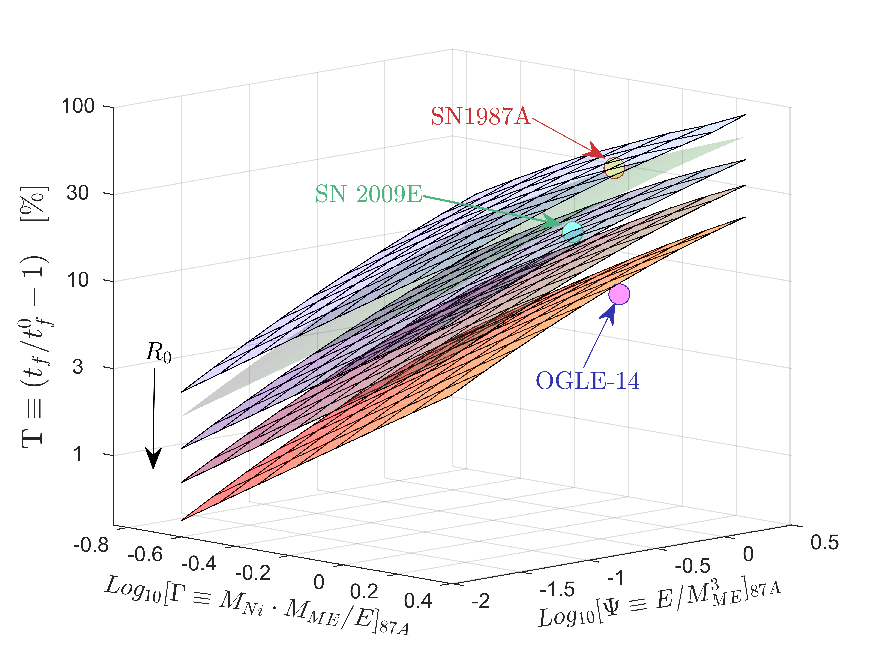} 
\caption{Relative time delay \textrm{T} in log-scale for 400 simulations inside the parameters space of typical 1987A-like events (see the text for details about the parameters' ranges). The different planes include all models with the same initial radius, and the grid-less one has
$[R_0]_{87A}=1$. The values obtained for the real well-observed objects SN 1987A, SN 2009E and OGLE-2014-SN-073 (OGLE-14, hereafter) are also reported for sake of comparison (see Table \ref{Tab:models} for the modeling parameters adopted for these real SNe).
\label{Fig:hyperplane_tf}}
\end{figure}

\begin{figure}
\includegraphics[angle=0,width=90mm]{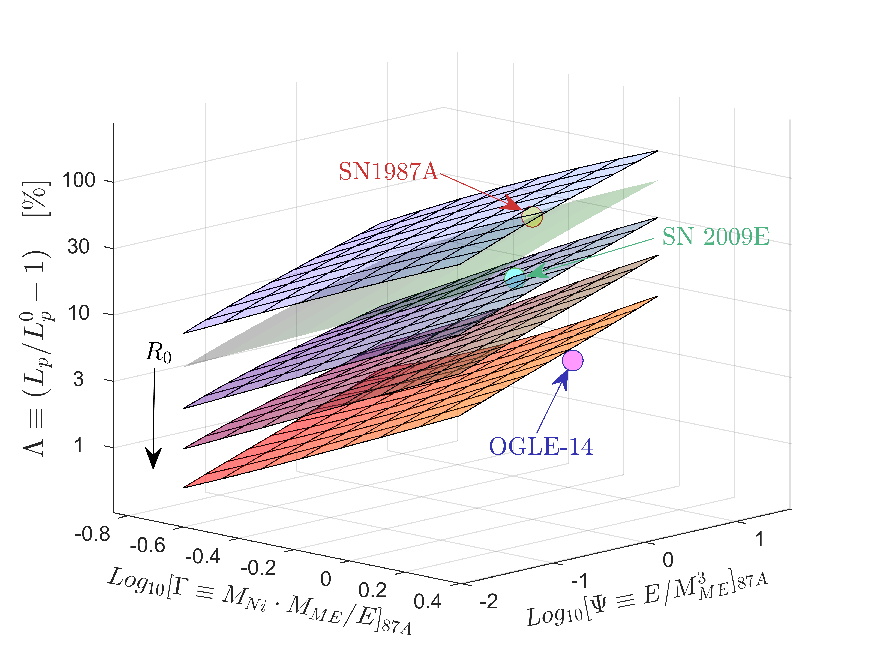} 
\caption{As Fig. \ref{Fig:hyperplane_tf}, but for the relative luminosity shift $\Lambda$.
\label{Fig:hyperplane_Lp}}
\end{figure}

As for $t_f'$ that, together with $L_p'$, represents the \chem{56}{Ni}-dependent parts of relations (\ref{Eq:Lt_2comp}), its behaviour depends only on the WCR evolution. Indeed $t_f'$ is essentially the difference between the recombination end time inferred from  equation (\ref{Eq:exp_diff}) and the one evaluated through the relation (\ref{Eq:Popov_eq15}). As a consequence, $t_f'$ can be described by only three parameters which, by analogy with the formalism used in equation (\ref{diff_eq_mat}), are $y_0$, $\lambda$, and $k_1$. The same is not valid for $L_p'$, which depends on four parameters, being linked to $L_{SN}$ (see Section \ref{SubSec:quarta_1}). However, since both $L_p^0$ and $L_p$ must have the same dependence on $L_a$ [cf.~equation (\ref{Eq:L_SN_y}) and note \ref{note_ti}], their ratio again dependent solely on the same three parameters affecting $t_f'$ (namely, $y_0$, $\lambda$, and $k_1$). Once individuated these parameters, in order to conduct a systematic analysis on the relative variation of the quantities $L_p$ and $t_p$ due to the \chem{56}{Ni} effects, it is useful to rewrite equations (\ref{Eq:Lt_2comp}) as follows:
\begin{equation}\label{Eq:Complete_scaling}
 \begin{cases}
 t_f=t_f^0(R_0,E,M_{ME})\times\left[1+\text{\textrm{T}}(y_0,\lambda,k_1)\right]\\
 L_{p}=L_p^0(R_0,E,M_{ME})\times [1+\Lambda(y_0,\lambda,k_1)];
 \end{cases}
 \end{equation}
where $\text{\textrm{T}}\equiv t_f/t_f^0-1=t_f'/t_f^0$ and $\Lambda\equiv L_p/L_p^0-1=L_p'/L_p^0$ are the relative time and luminosity corrections due to the \chem{56}{Ni} effects, respectively. Thus, \textrm{T} and $\Lambda$ depend only on the triplet of parameters $(y_0,\lambda,k_1)$, whose links to the modelling parameters are made explicit by relations (\ref{Eq:y_0})-(\ref{Eq:k_1}). From the latter relations, it is also possible to identify an ``orthogonal'' basis of parameters, defined by the relations 
\begin{equation}
\label{Eq:basis}
\Gamma\equiv\left[M_{Ni}\times M_{ME}/E\right],\quad \Psi\equiv \left[E/M_{ME}^3\right]\quad \text{and}\quad \text{\textRho} \equiv [R_0],
\end{equation} 
with which $y_0\propto \text{\textRho}^{1/2}\times\Psi^{1/4}$, $\lambda\propto\Gamma\times\Psi^{1/2}$ and $k_1\propto\Psi^{-1/4}$ span the entire modelling parameter space. The typical ranges of these orthogonal parameters for the class of 1987A-like SNe are:
\begin{equation}
\label{Eq:renges_basis}
\left[\Gamma\right]_{87A}\in[0.25,2.5]\quad\left[\Psi\right]_{87A}\in[0.01,1.5]\quad\left[\text{\textRho}\right]_{87A}\in[0.5,10];
\end{equation}
where the $[]_{87A}$ brackets indicate that the quantities reported inside them, are in units of SN 1987A's parameters (see Table 3 of \citetalias{pumo23} for typical ranges of parameters referred to 1987A-like SNe and Table \ref{Tab:models} for the HM parameters of SN 1987A). In this way, it is possible to numerically evaluate the \textrm{T} and $\Lambda$ corrections for synthetic SNe representative of SN 1987A-like events. Figs \ref{Fig:hyperplane_tf}-\ref{Fig:hyperplane_Lp} show these corrections for a grid of 400 synthetic SNe evaluated with our sub-model ``EXP+SOE'' and having modelling parameters uniformly distributed inside the parameters' ranges of equation (\ref{Eq:renges_basis}). This survey of simulations yields evidence of the direct multi-linear correlations between the logarithms of the relative corrections and the orthogonal parameters' ones, which are confirmed by the F-tests having determination coefficients $\gtrsim 0.99$ and p-values $\simeq 0$. Given this, it is then possible to write \textrm{T} and $\Lambda$ functions as follows
\begin{equation}
\label{Eq:linear_apx}
\begin{cases}
\text{\textrm{T}}=[\Gamma]_{87A}^{c_1}\times[\Psi]_{87A}^{c_2}\times [\text{\textRho}]_{87A}^{c_3}\times 10^{c_4}\\
\Lambda=[\Gamma]_{87A}^{c_5}\times[\Psi]_{87A}^{c_6}\times [\text{\textRho}]_{87A}^{c_7}\times 10^{c_8},
\end{cases}
\end{equation}
where all exponents $\{c_i\}$ (with $i= 1, ..., 8$) are the linear coefficients of the logarithmic forms for these equations. By using a multi-linear regression algorithm on the logarithms of equations (\ref{Eq:linear_apx}) \citep[such as the Iteratively Reweighted Least Squares Linear fit, see e.g.][]{street88}, the values of these exponents can be derived for both relative corrections, obtaining
\begin{equation}
\label{Eq:exponents}
\text{\textrm{T}}:
\begin{cases}
 c_1=0.902\pm 0.006\\
 c_2=0.350\pm 0.003\\
 c_3=-0.544\pm 0.004\\
 c_4=-0.408\pm 0.003
\end{cases}
\quad\Lambda:
\begin{cases}
 c_5=1.002\pm0.002\\
 c_6=0.233\pm0.001\\
 c_7=-0.901\pm0.003\\
 c_8=0.270\pm0.002,
\end{cases}
\end{equation}
which respectively present a root-mean-square deviation around the fits of about $0.04$ and $0.02$.\par

\begin{table}
\centering
\caption{Modeling parameters for SN 2009E, SN 1987A and OGLE-14 (see Table 3 of \citetalias{pumo23} and references therein for further details). Masses are in solar units, progenitor radius in $10^{12}$\,cm, and energy in foe ($\equiv$10$^{51}$\,ergs).}
\begin{tabular}{ccccc}
\hline\hline
SN & $M_{Ni}$           & $R_0$           & $E$   & $M_{ME}$          \\ 
   & [$\text{M}_\odot$] & [$10^{12}$\,cm] & [foe] & [$\text{M}_\odot$]\\    
\hline
SN 2009E & $0.04 $ & $7$  & $0.6$  & $19$ \\
SN 1987A & $0.075$ & $3$  & $1.3$  & $16$ \\
OGLE-14  & $0.47$  & $38$ & $12.4$ & $60$ \\
\hline
\end{tabular}
\label{Tab:models}
\end{table}


Using the relations (\ref{Eq:exponents}) and inserting equation (\ref{Eq:basis}) into (\ref{Eq:linear_apx}), it is possible to retrieve information about the dependence of \textrm{T} and $\Lambda$ by the modelling parameters. In particular, both \textrm{T} and $\Lambda$ depend on $M_{Ni}$ in an approximately linear way (given that $c_1\simeq c_5\simeq 0.9-1.0$), and decrease increasing $E$ (given that $c_2-c_1\simeq -0.55$ and $c_6-c_5\simeq -0.77$ are negative). This appear quite reasonable because the corrections have to be as great as the ratio between the radioactive source energy ($\propto M_{Ni}$) and the explosion one is greater. Moreover, differently from $t_f^0$ and $L_p^0$ that increase with $R_0$ [cf.~equation (\ref{Eq:Popov_scaling})], a greater value of $R_0$ reduces both \textrm{T} and $\Lambda$ (given that $c_3$ and $c_7$ are negative). This result demonstrates how the \chem{56}{Ni} effects become more important for SNe with small-radius progenitors, in agreement with the findings of \citetalias{pumo23}. Furthermore, as for $M_{ME}$, it produces opposite effects on \textrm{T} and $\Lambda$. Indeed, \textrm{T} decreases with the growth of $M_{ME}$ (given that $c_1-3c_2\simeq-0.15$ is negative), while $\Lambda$ grows up (given that $c_5-3c_6\simeq 0.31$ is positive).\par

Once evaluated how \textrm{T} and $\Lambda$ depend on the modelling parameters, it is also possible to write the following relations:
 \begin{equation}\label{Eq:scaling_00}
 \begin{cases}
 t_f'=t_f^0\times\text{\textrm{T}} \propto M_{Ni}^{0.90}\times R_0^{-0.37}\times E^{-0.72}\times M_{ME}^{0.35}\\
 L_p'=L_p^0\times\Lambda\propto M_{Ni}^{1.0}\times R_0^{-0.23}\times E^{0.06}\times M_{ME}^{-0.19},
 \end{cases}
 \end{equation}
that can be used to explain why the accuracy of scaling equations obtained from models neglecting the \chem{56}{Ni} effects like, in particular, the one of \citet[][]{popov93}, seems to be strongly dependent on the modelling parameters of the reference SN, as pointed out in \citetalias{pumo23}. Indeed, since the correction terms depend not only on the modelling parameter $M_{Ni}$ but also on the other parameters, if the reference SN's configuration at explosion is similar to that of the SNe to be carachterized through scaling relations, the values of \textrm{T} and $\Lambda$ are comparable and, consequently, the relations (\ref{Eq:Complete_scaling}) can be used as scaling equations in essence independent of $M_{Ni}$ and having the same dependencies on the modelling parameters of $t_f^0$ and $L_p^0$ [cf.~relations \ref{Eq:Popov_scaling}]. Conversely, if the reference SN significantly differs in terms of all modelling parameters to the other SNe, then their values of \textrm{T} and $\Lambda$ are no longer similar, and the equations (\ref{Eq:Complete_scaling}) can not be used as scaling relations. For example the latter is the case of OGLE-14, whose modelling parameters are widely different from those of the prototype of the class SN 1987A. This leads to have values of \textrm{T} and $\Lambda$ for OGLE-14 about one order of magnitude less than those for SN 1987A (cf.~Figs \ref{Fig:hyperplane_tf}-\ref{Fig:hyperplane_Lp}).\par

\begin{figure}
\includegraphics[angle=0,width=90mm]{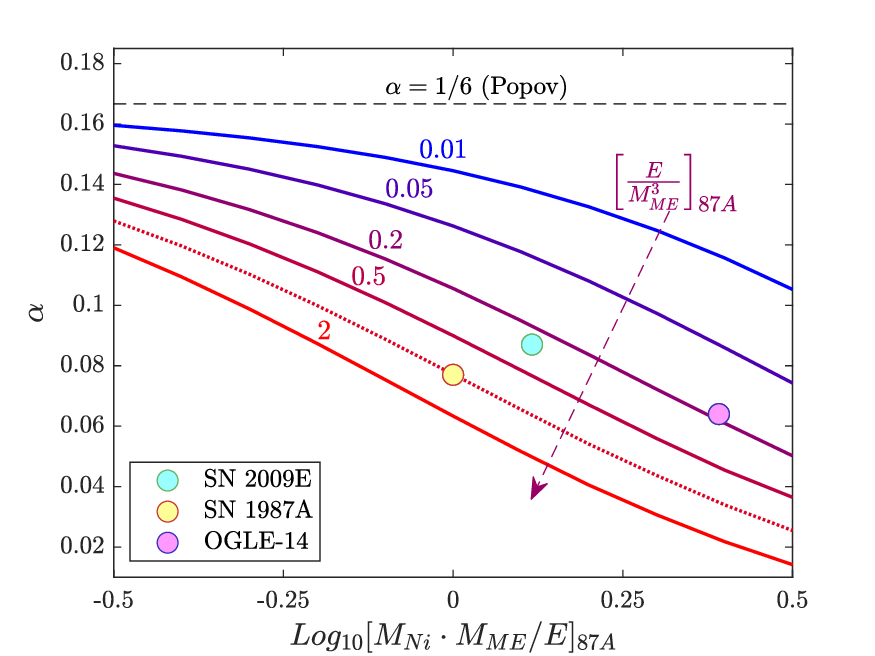} 
\caption{Behavior of $\alpha$ as a function of $\Gamma$ for different fixed values of $\Psi$ (see the different colored solid lines). The colored label above each solid line is the fixed value of $\Psi$. For sake of comparison,  the values of $\alpha$ obtained for the model of \citealt{popov93} (see the dotted line) and for the real well-observed objects SN 1987A, SN 2009E and OGLE-14 (see the colored circles) are also reported (see Table \ref{Tab:models} for the modeling parameters adopted for these real SNe).  
\label{Fig:alfa_exp}}
\end{figure}

The deviation of equations (\ref{Eq:Complete_scaling}) from being simple scaling equations becomes more evident when focusing on the dependence of $t_f$ on $R_0$, that can be expressed by a power-law relation of index $\alpha$. For models neglecting the \chem{56}{Ni} effects, the $\alpha$ exponent is a constant [in particular equal to $1/6$ for the model of \citet{popov93}; see also relations (\ref{Eq:Popov_scaling})]. For models including the \chem{56}{Ni} effects, the Ni introduction modifies this simple dependency, making $\alpha$ not constant. In particular, for our new model, $\alpha$ is a function of $\Gamma$ and $\Psi$ (see Fig.~\ref{Fig:alfa_exp}), so its value depends on $M_{Ni}$, $E$, and $M_{ME}$. Specifically, for SNe with similar scale velocity ($v_{sc}^2\propto E/M_{ME}=cost$), $\alpha$ decreases when increasing $M_{Ni}$, especially for SNe having less massive ejecta. Note that the parameters of SN 1987A give $\alpha\simeq 0.08$, which is about the half of what is expected by the model of \citet{popov93}. Similarly, also SN 2009E and OGLE-14, with $[\Psi]_{87A}$ around $0.2$ for both and $[\Gamma]_{87A}$ respectively equal to $0.13$ and $0.40$, give $\alpha$-values between $0.07$ and $0.09$ (cf.~Fig.~\ref{Fig:alfa_exp}).\par

In general, although the equations (\ref{Eq:Complete_scaling}) can not be directly used as scaling relations, they could be adopted as ``exact'' equations, that can be rewritten as:
\begin{equation}
\label{Eq:system}
\begin{cases}
t_f= 17.2\>\text{days}\times R_0^{1/6}\> E^{-1/6}\> M_{ME}^{1/2}\times [1+\textrm{T}]\\
L_p= 5.40\times 10^{41}\> \text{ergs/s}\times R_0^{2/3}\> E^{5/6}\> M_{ME}^{-1/2}\times [1+\Lambda],
\end{cases}
\end{equation}
where
\begin{equation}
\begin{cases}
\textrm{T}= 12.59\times M_{Ni}^{0.90}\times R_0^{-0.54}\times E^{-0.55}\times M_{ME}^{-0.15}\\
\Lambda= 33.19\times M_{Ni}^{1.0}\times R_0^{-0.90}\times E^{-0.77}\times M_{ME}^{0.31},
\end{cases}
\end{equation}
in which all explosive parameters are in the same units of Table \ref{Tab:models}.\par

\begin{figure}
\includegraphics[angle=0,width=92mm]{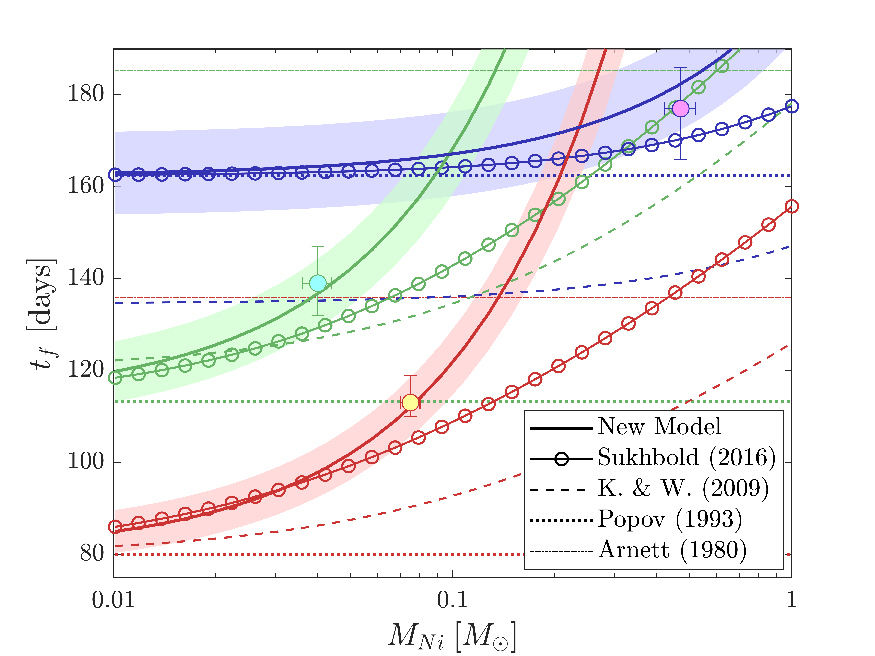} 
\caption{Dependence of $t_f$ on the modelling parameter $M_{Ni}$ for the three different fixed triplets of the other modelling parameters ($R_0, E, M_{ME}$) reported in Table \ref{Tab:models}, corresponding to SN 2009E (green solid line), SN 1987A (red solid line), and OGLE-14 (blue solid line). The values of $t_f$ are evaluated using the relation (\ref{Eq:Complete_scaling}) derived from our new model. Shaded areas are the error regions when assuming an uncertainty on $E$, $M_{ME}$, and $R_0$ of $10\%$. The values of $t_f$ deduced from the observations of SN 2009E (cyan circle with green perimeter), SN 1987A (yellow circle with red perimeter), and OGLE-14 (purple circle with blue perimeter) are also reported. In particular, for each real SN, the reported value is the epoch of the first observation in which its magnitude error is compatible with the radioactive tail extension, while the error on $t_f$ is evaluated using the epochs of the closest observations (see also Section \ref{SubSec:quarta_3} for further details). For sake of comparison, the curves showing the recombination end time evaluated according to the models presented in \citet{arnett80}, \citet{popov93}, \citet{KW09}, and \citet{sukhbold16} are also reported. Likewise $t_f$, it is reported the dependence on $M_{Ni}$ for the three above mentioned fixed triplets ($R_0, E, M_{ME}$).
\label{Fig:tf_relation}}
\end{figure}

\begin{figure}
\includegraphics[angle=0,width=92mm]{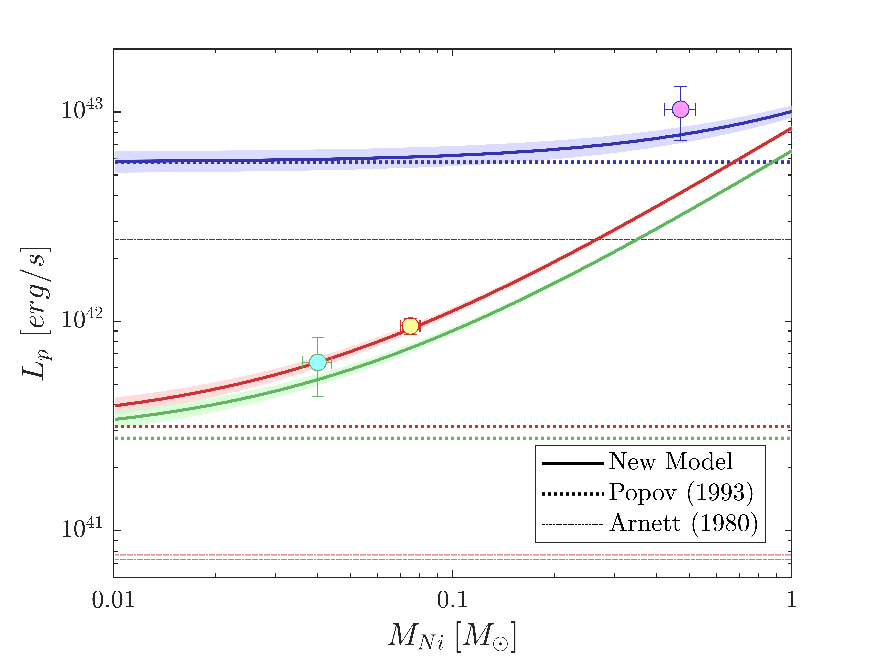} 
\caption{As Fig.~\ref{Fig:tf_relation}, but for $L_p$. For each real SN, the reported value is the highest luminosity observed during the peak stage. As for the curves showing the $L_p$ behavior according to other  models present in literature, those evaluated according to the models presented in \citet{KW09} and \citet{sukhbold16} are not shown because they are equal to that evaluated according to the model of \citet{popov93}.}
\label{Fig:Lpeak_relation}
\end{figure}

The modeling parameters deduced from the HM of SN 1987A, SN 2009E and OGLE-14 (see Table \ref{Tab:models}), allow us to test these new analytical relations on real SNe and to compare them with other similar relationships reported in the literature (see Figure \ref{Fig:tf_relation}-\ref{Fig:Lpeak_relation}).\par


Concerning the $t_f$ prediction (see in particular Fig.~\ref{Fig:tf_relation}), the diffusive equation proposed by \citet{arnett80} overestimates its value for all three SNe by about 20-40 days whereas, as expected for not Ni-free SNe \citep[see e.g.~also][]{PZ13}, the analytic expression presented in \citet{popov93} systematically tends to underestimate the recombination end time. Focusing on the parameters of SN 1987A, the relations proposed by \citet{KW09} and \citet{sukhbold16} lead to a recombination end time, respectively, about 0.6 and 0.9 times lower than the value deduced using our new relation. The relatively low value obtained with the relation of \citet{KW09} seems primarily attributable to the different dependence of the recombination end time on $E$. Indeed \citet{KW09} adopt a power-law relation with index $-1/4$, differently from both our relation and the relation presented in \citet{sukhbold16} where the index is $-1/6$. Instead, the discrepancy between our value and the one found using the relation of \citet{sukhbold16} appears mainly attributable to the different dependence of the recombination end time on $M_{Ni}$. Indeed, compared with the relation presented by \citet{sukhbold16}, the first equation of set (\ref{Eq:system}) is more sensible to the $M_{Ni}$ variation, while the other parameters dependencies follow a quite similar trend. Furthermore, our new relation provides the best agreement with the observations, whose deviation of few days not only are inside the errors but also expected because set (\ref{Eq:system}) has been derived using the sub-model ``EXP+SOE'' (which tends to slightly underestimate $t_f$; cf.~Section \ref{Sec:terza}).\par

Concerning the $L_p$ prediction (see also Fig.~\ref{Fig:Lpeak_relation}), differently from the other relations present in literature, our relationship depends also on $M_{Ni}$. The \chem{56}{Ni} effects are particularly important in cases such as SN 1987A and SN 2009E, where the luminosity reaches approximately three times the value predicted by the Popov's model \citep{popov93}, which in turn is about twice the value predicted by the relation of \cite{arnett80}. We also remark that, according to our relation, the link between $L_p$ and $M_{Ni}$ (and, more in general, between the LC peak features and $M_{Ni}$) cannot be described by a simple proportionality relation, at least until \textrm{T} (and, more in general, also $\Lambda$) is sufficiently large (i.e.~$\gtrsim 1$). Therefore the shape of the peak depend on $M_{Ni}$ according to a ``threshold'' behaviour, which emerges only when the value of $M_{Ni}$ is sufficiently high (typically greater than $10^{-2}\,\text{M}_\odot$), confirming what found in \citetalias{pumo23} using an independent approach.\par

Last but not least note that, given the broad applicability of our new model, which is in principle pertinent to any types of H-rich SNe, the new original relationships (\ref{Eq:system}) can be applied to the plateau phase of type IIP SNe. Additionally, these relationships offers a way to verify the theoretical consistency about the local and total gamma-ray thermalization's assumptions. Specifically, using the first equation of system (\ref{Eq:system}), the ratio between the gamma-ray mean free path within the ejecta $l_\gamma(t)$ [$\equiv 1/k_\gamma \rho(t)$] and the ejecta radius at the end of the recombination phase can be expressed as:
\begin{equation}
\frac{l_\gamma(t_f)}{R(t_f)}= \frac{(t_f+t_e)^2}{k_\gamma \rho_0 t_e^3 v_{sc}}
\simeq 25.5\% \times \left[R_0^{1/3} E^{2/3} M_{ME}^{-1} \times (1+\textrm{T})^2\right],
\end{equation}
where $k_\gamma\simeq 0.033$ cm$^2\,$g$^{-1}$ is the average gamma-rays opacity \citep[][and references therein]{balberg00}. Using the explosion parameters from Table \ref{Tab:models}, we find that the highest value of this ratio is for OGLE-14, approximately 11\%. For SN1987A and SN2009E, the ratio respectively decreases to 5\% and 2\%, and for a typical Type IIP SN with a radius at explosion about ten times larger than that of SN1987A, the ratio is estimated to be around 6\%. These findings suggest that, in scenarios where more than 99\% of $M_{Ni}$ is confined within 60\% of the ejecta's radius (as in our cases ``EXP'' or ``BOX''), the gamma-ray mean free path remains smaller than the minimum distance required for gamma rays to escape the ejecta up to a time $\sim 2t_f$ at least. Consequently, for SN 1987A and other similarly massive H-rich SNe (with ejecta masses $\gtrsim 10$\msun), it is reasonable to assume total thermalization of gamma rays produced by \chem{56}{Co} decay, even during late phases as already noticed in \citet[][]{zampieri17}. As for the non-local thermalization effects, it is possible that gamma rays produced above the WCR are not fully thermalized within the transparent region and, consequently, contribute to heating the inner opaque reagion. This represents a third-order effect in the description of the Ni-source energy contributions in the sources functions $S_i$ and $S$, which in turn can slightly influence the behavior of LC, especially toward the end of the recombination phase. Indeed, this effect could alter the evolution of the WCR only when the gamma-ray mean free path becomes comparable to the radius of the recombination front (i.e.~when $x_i\sim \l_\gamma/R\lesssim 0.1$). For 1987A-like events, this condition is met only a few days before $t_f$ (see also Fig.~\ref{Fig:xiprofiles}), making it a minor contribution compared to the differences arising from the assumptions in the ``SOE'' and ``IE'' cases. Consequently, the impact of non-local effects is negligible and falls within the current observational error bars. However, a more detailed quantification of these effects would require the development of a model that explicitly addresses the problem of energy transport for gamma rays within the ejecta \citep[see e.g.][]{balberg00,KK19}, which is outside the scope of this paper.\par

\subsection{LC features of SN 1987A \& Scaling relations}\label{SubSec:quarta_3}

In addition to the peak features, the LC of SNe resembling SN 1987A usually exhibit additional observational features that can be linked to some parameters describing the SN progenitor at explosion through simple scaling equations [see e.g.~relation (13) of \citetalias{pumo23}]. Unlike the relations for $L_p$ and $t_f$ presented in Section \ref{SubSec:quarta_2}, the scaling relations have to be characterized by a direct proportionality between these parameters and measurable photometric (or, more in general, spectrophotometric) quantities. So, in order to identify common features of the SN 1987A-like objects that can be correlated to the main parameters describing the SN progenitor at explosion through ``pure'' photometric scaling relations based on our new model, we closely examine the absolute magnitude evolution of the prototype of the class SN 1987A (i.e.~$M_{bol}\equiv -2.5\,\log_{10}L_{Obs.}+88.7$) and its temporal derivative\footnote{To describe the temporal derivative of $M_{bol}$, the observed LC of SN 1987A has made continuous and derivable through a interpolation process based on the Gaussian Process (GPR, heheafter). In particular, following \citet[][and references therein]{Inserra18}, we use a GPR characterized by a constant basis function and the Matern-32 kernel.\label{Note:GPR}} ($\dot{M}_{bol}$, hereafter).\par
\begin{figure}
  \includegraphics[angle=0,width=92mm]{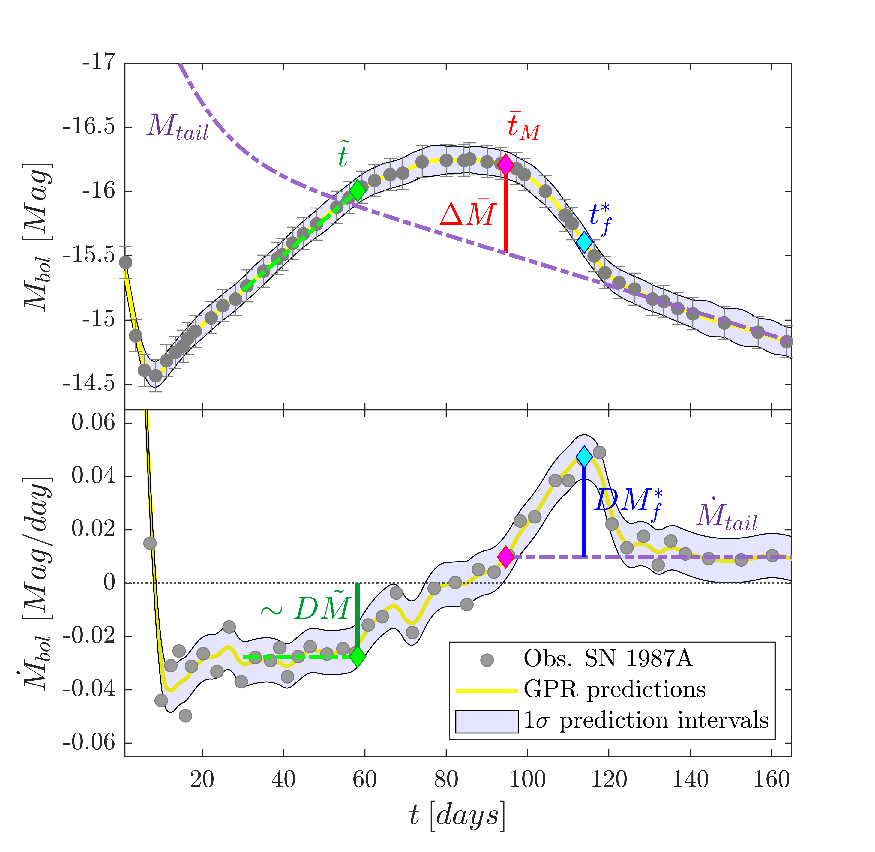} 
  \caption{Bolometric magnitude of SN 1987A (top panel) and its time derivative (bottom panel) as a fuction of time. The quantities $\tilde{t},\>\bar{t}_M,\>t_f^*,\>D\tilde{M},\>\Delta \bar{M}$, and $DM_f^*$ (see text for details) have been evaluated after having applied a GPR-based precedure (see note \ref{Note:GPR} for details). \label{Fig:LC_features}}
 \end{figure}
 
We find that, during the rising phase, the $M_{bol}$ behavior appears approximately linear until a certain epoch indicated as $\tilde{t}$ (see the green diamonds of Fig.~\ref{Fig:LC_features}). In particular, this phase is characterized by an almost constant $\dot{M}_{bol}$ value (roughly equals to $-0.03$ for SN 1987A; see the green dashed line in bottom panel of Fig.~\ref{Fig:LC_features}) between 30 and $\tilde{t}$ (roughly equals to 60 for SN 1987A) days after the explosion. This observed behavior is reproduced by our new model that includes the \chem{56}{Ni} effects. In particular, using our new model, the derivative of the magnitude in the time interval $[30$ d, $\tilde{t}]$ (i.e.~the rate of the magnitude in the rising phase ranging from to 30 d to $\tilde{t}$, denoted as $\dot{M}_{rising}$ hereafter) can be approximated by the expression (see Appendix \ref{Appendix C} for details about its demonstration and related assumptions)
\begin{equation}
\label{Eq:risingM_simp}
 \dot{M}_{rising}(t) \simeq-(2.5/\ln{10})\times\left[\frac{1}{t}+\frac{M_{Ni}\>\epsilon_{^{56}Co}\>\tau_{^{56}Co}}{L_a\>t_i\>\left(\tau_{^{56}Co}-\tau_{^{56}Ni}\right)}\right],
\end{equation}
showing that $\dot{M}_{rising}$ is dependent on two terms. The first one is inversely proportional to time and independent of the modelling parameters \citep[like in the models non considering \chem{56}{Ni} effects, see e.g.][]{popov93}. The second term, not present in models neglecting the \chem{56}{Ni} effects, is time-independent and directly proportional to $M_{Ni}$. Just the presence of this second term linked to the \chem{56}{Ni} effects and, in particular, its balance with the first one explains why the rising phase of secondary peak is linear up to $\tilde{t}$. Indeed, for $t>\tilde{t}$, the first (and the other neglected terms, see Appendix \ref{Appendix C}), which represents the ejecta cooling due to the expansion, makes $\dot{M}_{rising}$ not any more linear. Moreover, adopting the observed rate of the magnitude at $\tilde{t}$ [namely, $\dot{M}_{bol}(\tilde{t})$] as a measure of $\dot{M}_{rising}(\tilde{t})$, and defining the quantity $D\tilde{M}$ as
\begin{equation}
\label{Eq:DM_tilde}
D\tilde{M}\equiv \frac{2.5\,\log_{10}{e}}{\tilde{t}}+\dot{M}_{rising}(\tilde{t}),
\end{equation}
it is possible to correlate the value of $D\tilde{M}$ inferred from the observations to the main parameters describing the SN progenitor at explosion, through the following photometric scaling relation based on our new model [see Appendix \ref{Appendix C} and, in particular, equation (\ref{Eq:scaling1_app}) for further details]:
\begin{equation}
\label{Eq:scaling1_text}
D\tilde{M} \propto \frac{M_{Ni}}{L_a\,t_i}\propto M_{Ni}\times R_0^{-0.5}\times E^{-0.5}\times M_{ME}^{-0.5}.
\end{equation}
Furthermore, in a reverse approach, this scaling equation enables us to more deeply understand the LC behavior during the rising phase in SN 1987A-like events. In particular, as expected, the rise in brightness increases in proportion to $M_{Ni}$, because the \chem{56}{Ni} decay provides an additional amount of energy compared to $E$. The increasing of $E$, on the other hand, tends to dampen the LC rising, since a greater value of $E$ corresponds to a faster expansion rate and, consequently, a faster cooling rate of the ejecta. Moreover, the increasing of $M_{ME}$ leads to increase the scattering time of photons, postponing the loss of energy due to electromagnetic emission. So, when increasing $M_{ME}$, the luminosity tends to grow over the time more slowly. Also the value of $R_0$ has an impact on the rising phase, explaining why SNe with radii at explosion 10-100 times larger than those of 1987A-like SNe tend to exhibit a plateau instead of a secondary peak \citep[see e.g.~also][and reference therein]{PZ11}. This outcome confirms the connection between compact progenitors and 1987A-like SNe. The only exception is the high-mass ($\gtrsim 30$\,\Msun) and high-energy ($\gtrsim 10$\,foe) tail of SN 1987A-like events, linked to extended progenitors with values of $R_0\sim 10 ^{13}$-$10^{14}$\,cm, that also challenges standard theories of neutrino-driven core-collapse and stellar evolution (see \citetalias{pumo23} for further details).\par

In order to find other observational features that can be used when applying scaling equations to real SNe, we continue the LC analysis focusing on subsequent epochs, like the maximum of the secondary peak and the post-peak phases. In this context, it is necessary to be able to evaluate the following quantity $M_{tail}$ from the observational data:
\begin{equation}
\label{Eq:M_tail}
M_{tail}(t)=-2.5\log_{10}\left[\epsilon(t)\right]+m_{Ni},
\end{equation}
where $m_{Ni}$ is a constant linked to the modelling parameter $M_{Ni}$, but which can be inferred from the observations by fitting $M_{bol}$ during the radioactive tail. Once evaluated $M_{tail}$, it is also possible to estimate the difference $M_{bol}-M_{tail}$. By defining $\bar{t}_M$ as the epoch when this difference is maximum\footnote{$\bar{t}_M$ is also the abscissa in which the derivatives of $M_{bol}-M_{tail}$ and $\Delta M$ are null (see Fig.~\ref{Fig:LC_features}). Therefore, it results $\dot{M}_{bol}(\bar{t}_M)=\dot{M}_{tail}(\bar{t}_M)\simeq -2.5\,\tau_{^{56}Co}^{-1}\,\log_{10}e$ [cf.~equation (\ref{Def:DM_f_star})].\label{Note:t_barM}} (see the purple diamonds and the vertical red line of Fig.~\ref{Fig:LC_features}), the quantity
\begin{equation}
\label{Def:DeltaMbar}
\Delta \bar{M}\equiv M_{bol}(\bar{t}_M)-M_{tail}(\bar{t}_M)
\end{equation}
can be used to evaluate the luminosity $\bar{L}_M= 10^{-\Delta \bar{M}/2.5}$, which is involved in the further photometric scaling relation based on our new model reported in the following [see Appendix \ref{Appendix C} and, in particular, equation (\ref{Eq:scaling2_app}) for further details]:
\begin{equation}
\label{Eq:scaling2_text}
\bar{t}_M^8\,(\bar{L}_M-1)^{-2} \propto M_{Ni}^2\times E^{-3}\times M_{ME}^{5}.
\end{equation}
This relationship is not a merely correlation between photometric features at maximum and parameters describing the SN progenitor at explosion. Indeed, it is necessary to evaluate $\bar{t}_M$ and $\Delta \bar{M}$ that, in turn, implies to know the LC behaviour also after the rise to the maximum up to the beginning of the radioactive tail. This is in agreement with the findings of Section \ref{SubSec:quarta_2}, according to which it is not possible to establish a direct proportionality between photometric features at maximum like $L_p$ and the modelling parameters [cf.~relations (\ref{Eq:system})].\par

Another photometric feature of SN 1987A-like events that can be linked to some parameters describing the SN progenitor at explosion through a scaling equation, concerns the LC behaviour just before the beginning of the radioactive tail. In particular, analysing the LC and its derivative as a function of time, it is possible to notice a sudden change in the value of $\dot{M}_{bol}$ at this phase  (see Fig.~\ref{Fig:LC_features}) which, from a physical point of view, corresponds to the final phase of the ejecta recombination. A similar behavior is also reproduced by our model, that shows a jump exactly at time $t_f$ [see Appendix \ref{Appendix C} and, in particular, equation (\ref{Eq:D_M}) for further details]. However, for real SNe, the behavior is smoother and, consequently, the inflection point $t_f^*$, which corresponds to a local maximum of $\dot{M}_{bol}$ (see the cyan diamonds of Fig.~\ref{Fig:LC_features}), does not precisely coincide with $t_f$, slightly preceding it. The physical reason for this difference could be linked to the ejecta density profile of real SNe, whose inner zones may not be well reproduced by the uniform density assumption used to describe the ME in our model. To correctly quantify this second-order effect linked to the density profile, it should be necessary to develop a (semi-)analytic model considering density variations inside the ME, which is outside the scope of this work. Nevertheless, in the case of SN 1987A, the difference $t_f-t_f^*$ is about 5 d, that is about $5\%$ of the value of $t_f$. This makes the value of $t_f^*$ still compatible with the one of $t_f$ obtained from equation (\ref{Eq:system}), where we assume a uncertainty on $t_f$ of 10\% (see Fig.~\ref{Fig:tf_relation}). So the value of $t_f^*$ can be adopted as a measure of $t_f$. Consequently $t_f^*$ can be also used in the following photometric scaling relation based on our new model [see Appendix \ref{Appendix C} and, in particular, equation (\ref{Eq:scaling3_app}) for further details] and simplified for SNe with similar\footnote{When comparing two SNe with a similar $t_f^*$ using the scaling equation (\ref{Eq:scaling3_app}), the exponential factor $e^{-t_f/\tau}$ present in  the equation can be simplified. Indeed, considering two different SNe with $t_1$ and $t_2$ as $t_f^*$, the ratio between the exponential factors $e^{-(t_1-t_2)/\tau}$ is $\simeq 1$ when $|t_1-t_2|<<\tau$.} $t_f$:
\begin{align}
\label{Eq:scaling3_text}
[t_f^*]^6 \,[DM_f^*]^{-2}\propto M_{Ni}^2\times E^{-3}\times M_{ME}^{5},
\end{align}
in which $DM_f^*$, adopted as an estimation of $DM_f$ [cf.~equation (\ref{Eq:scaling3_app})], is  defined as
\begin{equation}
\label{Def:DM_f_star}
DM_f^*\equiv \dot{M}_{bol}(t_f^*)-\dot{M}_{tail}(t_f^*)\simeq\left.\frac{dM_{bol}}{dt}\right|_{t_f^*}-2.5\,\tau_{^{56}Co}^{-1}\,\log_{10}e,
\end{equation}
where $\dot{M}_{bol}({t_f^*})$ is the time derivative of $M_{bol}$ at the inflection epoch $t_f^*$, and $\dot{M}_{tail}({t_f^*})$ is the time derivative of $M_{tail}$ at $t_f^*$ evaluated by means of the relation (\ref{Eq:decay_approx}). Compared to the scaling relation (\ref{Eq:scaling2_text}), the relation (\ref{Eq:scaling3_text}) have the advantage of directly linking the modelling parameters with only late-time LC features. However, it depends on the LC derivative, which requires to well sample the LC also long after the peak. Considering the modelling parameters dependencies in the scaling equation (\ref{Eq:scaling3_text}), they are the same as those obtained for equation (\ref{Eq:scaling2_text}). This makes the two realtions completely equivalent for the purpose of inferring information about the main parameters describing the SN progenitor at the explosion, testifying that both the characteristics of the peak and the recombination ending are related by the same combination of modelling parameters. In particular, as expected, the increase in $M_{Ni}$ brings a larger amount of heat into the ejecta, which therefore takes longer to cool and recombine, thus also reducing the rate at which the magnitude decreases (note that $\bar{L}_M$ and $DM_f^*$ are $\propto M_{Ni}^{-1}$). $M_{ME}$ has an even more important overall effect on the recombination ending, that precisely derives from its link with the diffusion time (namely, $t_f^*\propto M_{ME}^{5/6}$). Finally, the dependence on $E$ shows that the cooling process is as fast as the ejecta expands faster, thus anticipating both the peak phase and the recombination ending (note that $\bar{t}_M$ is $\propto E^{-3/8}$ and $t_f^*$ is  $\propto E^{-1/2}$).

\section{Summary \& Further Comments}\label{Sec:quinta}

With the intent of better understanding the SN 1987A-like events and, more in general, the H-rich SNe, we have developed a new analytic model, which is able to describe their entire post-explosive evolution (i.e.~from the breakout of the shock wave at the stellar surface up to the nebular stage). The distinctive features of the new model are the possibility to evaluate the bolometric emitted luminosity and the SN expansion velocity, taking into account the following three different effects: 1) the recombination of the ejected material, 2) the heating effects due to the \chem{56}{Ni} decay in the computation of the WCR evolution, and 3) the possible presence of an OTS not-homologously expanding and surrounding the bulk of the ejecta that, instead, is in homologous expansion. The second property represents the major novelty of this model. Indeed, differently from other models present in literature \citep[e.g.][]{arnett80,popov93}, our model includes $M_{Ni}$ among the main modelling parameters (together with $E$, $M_{ME}$, and $R_0$) and, in order to evaluate the heating effects due to the \chem{56}{Ni} decay through source terms, it is possible to consider both various spatial distributions of $^{56}$Ni and different angular emission distributions of the thermalized radiation due to the $^{56}$Ni decay. Moreover, thanks to the third property, it is also possible to consider (or not) the effects linked to the OTS presence, as well as to apply (or not) a correction linked to the position of the line formation region when evaluating the SN expansion velocity. In this way, one have various sub-models characterized by different accuracy, computing time demands, and capability of accurately reproducing the LC behavior and the SN expansion velocity during the three main post-explosive phases \citep[i.e.~diffusive, recombination, and radioactive-decay phases; see e.g.][for further details]{PZ11}.\par

So our new model can be used to different aims, ranging from the computation of huge grids of synthetic LCs and SN expansion velocities up to the accurate model fitting of single real events, passing through a sufficient realistic analytic description of the SN 1987A-like events (and, more in general, of H-rich SNe) suitable for studying their physical properties and linking the latter ones to their observational features. In this paper, in addition to present our new model, we use it to deeply analyse the link between photometric features of SN 1987A-like objects and parameters describing their progenitor's physical properties at the explosion, investigating also the modeling degeneration problem in H-rich SNe and the possibility to ``standardize'' the subgroup of the 1987A-like SNe.\par

As for the modelling degeneration problem, we find that its occurrence depends on the value of the free coefficient $\lambda$ defined by the relation (\ref{Eq:lambda}) and linked to the main SN progenitor's physical properties at the explosion. In particular, for values of $\lambda$ sufficiently low (i.e. $\lambda \lesssim 10^{-2}-10^{-3})$, it is possible  to reproduce essentially the same LC with more than one set of modelling parameters. Therefore, degeneration problems can arise when modelling the sole LC. In this case, the additional information of spectroscopic nature, such as that inferred by modelling the expansion velocity, has to be also used to uniquely constrain the set of modelling parameters. On the other hand, for SNe characterized by higher values of $\lambda$ like the real SN 1987A-like events, no significant degeneration problems arise when performing their LC modelling. Thus the additional information of spectroscopic nature could be used to constrain all the modelling parameters describing a real 1987A-like SN and, consequently, to characterize the event also if its distance (or equivalently its absolute bolometric luminosity) is not known. A further corollary of these findings may be the possibility of standardizing the 1987A-like SNe using spectrophotometric information, in accordance with what found in \citet[][]{PZ13}, where a purely photometric based standardization appears difficult to be realized for this subgroup of H-rich SNe.\par

Concerning the LC behavior of SN 1987A-like objects and, more in general, its link with the progenitor's physical properties at the explosion, we deduce two new Ni-dependent relationships, based on our model, which link some features of the 1987A-like SNe LC (namely, the peak luminosity and its width) to the values of the main modelling parameters (i.e.~$E$, $M_{ME}$, $R_0$, and $M_{Ni}$). Contrary to similar relations proposed in other works \citep[e.g.][]{arnett80,popov93,KW09,sukhbold16}, our new relationships are in excellent agreement with observations of real SNe and, given the wide feasibility of our new model, which in principle is usable for any types of H-rich SNe, they can be also applied to type IIP SNe (provided that the plateau luminosity and its duration replace, respectively, the peak luminosity and its width). However we remark that, as a rule (there could be some exceptions; cf.~Section \ref{SubSec:quarta_3}), our new relations cannot be used as scaling equations essentially because the $M_{Ni}$ effects on the LC peak features cannot be described by simple proportionality relations. Nevertheless, from our model it is possible to derive scaling relations. In particular, in this paper we present three new scaling relations, which may be used for estimating the main SN progenitor's physical properties, once only the photometric behaviour of the 1987A-like object is known.\par

In the light of these results, at present we are working to further check the validity of the new Ni-dependent relationships and scaling relations, directly analysing their robustness on an observed sample of 1987A-like SNe with sufficiently good photometric coverage. Moreover, we are further developing our new analytic model, so as to make it usable for other types of SNe (e.g.~interacting SNe) and other electromagnetic transients similar to SNe (e.g.~kilonovae). Indeed, the new mathematical formulation of the recombination problem adopted in our model [cf.~equation~(\ref{Eq:generalxi})], lets to change the type of sourcing mechanism with an appropriate choice of the source function \citep[see e.g.~also][]{matsumoto25}. This, in principle, gives us the possibility to extend our new model to any kind of electromagnetic transient, in which the presence of an internal source, such as a black hole or a magnetar, can significantly affect its LC \citep[see e.g.][and references therein]{DK13,KK19,moriya22}.

\section*{Acknowledgments}
This paper is supported by the Fondazione  ICSC, Spoke 3 Astrophysics and Cosmos Observations. National Recovery and Resilience Plan (Piano Nazionale di Ripresa e Resilienza, PNRR) Project ID CN\_00000013 ``Italian Research Center on  High-Performance Computing, Big Data and Quantum Computing'' funded by MUR Missione 4 Componente 2 Investimento 1.4: Potenziamento strutture di ricerca e creazione di ``campioni nazionali di R\&S (M4C2-19)'' - Next Generation EU (NGEU).
\section*{Data availability}
The data underlying this article are available in the article.

\bibliographystyle{aa}

\appendix
\section{Temperature profile and luminosity evolution}\label{Appendix A}

In this appendix we show in detail how, once defined the temperature through the relations (\ref{Eq:tempprof}) and (\ref{Eq:psi_t}), it is possible to evaluate $E_{th}^0$ by calculating the integral in the first of the relations (\ref{Eq:Eth_td}), derive relations (\ref{Eq:Lop}) and (\ref{Eq:generalxi}), and determine the time evolution of $L_{op}$ during the diffusive phase [i.e.~infer the relation (\ref{Eq:Lop_complete}) for $t<t_i$].\par

As for $E_{th}^0$, inserting the first of the equations (\ref{Eq:radtransport}) in the first of the relations (\ref{Eq:Eth_td}) and considering the relations (\ref{Eq:tempprof}) and (\ref{Eq:psi_t}), one obtains
\begin{align}
\label{Eq:Eth_ini}
E_{th}^0=\,& \bigintsss_0^{M_{ME}}\frac{aT^4}{\rho}\bigg|_{t=0}\> dm= 3aM_{ME}\bigintsss_0^{1}\frac{T^4(x,0)}{\rho_0}\> x^2dx= \nonumber\\
=\,& 4aR_0^3T_0^4\int_0^1 sin(\pi x)xdx= \frac{4aT_0^4R_0^3}{\pi},
\end{align}
where $dm=M_{ME}dx^3=3M_{ME}x^2dx$ [cf.~relation (\ref{Eq:mass})].\par

To derive relation (\ref{Eq:Lop}), it is first necessary to substitute the relations (\ref{Eq:tempprof}) and (\ref{Eq:psi_t}) into the last of the equations (\ref{Eq:radtransport}). Doing so and considering also the last of the relations (\ref{Eq:Eth_td}) and the relation (\ref{Eq:Eth_ini}) to simplify the notation, the outgoing luminosity from a shell $L(x,t)$ inside the opaque region (i.e.~$x\leq x_i$) can be written as
\begin{align}
\label{Eq:lum_strato}
L(x,t)=\,&-\frac{4\pi a c}{3k\rho_0 R_0^3}\> R^4(t)\> x^2\> \frac{\partial T^4(x,t)}{\partial x}= \nonumber\\
=\,&-\frac{4\pi^3cR_0}{9kM_{ME}}\times \frac{4 aT_0^4R_0^3}{\pi}\> \phi(t)\> x^2 \frac{\partial \psi_t(x)}{\partial x}= \nonumber\\
=\,&\frac{E_{th}^0}{t_d}\> \phi(t)\> x \> \left(\frac{\sin{\omega}}{\omega}-\cos{\omega}\right),
\end{align} 
where $\omega=\omega(x,t)\equiv\pi x/x_i(t)$. Then, considering that $\omega[x=x_i(t),t]=\pi$ and given that the relation (\ref{Eq:lum_strato}) evaluated for $x=x_i$ gives the outgoing luminosity from WCR $L_{op}$, one obtains
\begin{align}
\label{Eq:Lop_dim_App}
L_{op}(t)=\,&L[x=x_i(t),t]= \frac{E_{th}^0}{t_d}\> \phi(t)\> x_i(t)\> [0-(-1)]= \nonumber\\
=\,&\frac{E_{th}^0}{t_d}\> \phi(t)\> x_i(t) \equiv \text{relation (\ref{Eq:Lop}) (Q.E.D.)}.  
\end{align}

To derive the relation (\ref{Eq:Lop_complete}) for $t<t_i$, it is necessary to progress into three steps: in the first step $L_{op}$ is evaluated for $t<t_i$ using a different and independent method compared to that adopted for deriving relation (\ref{Eq:Lop}), in the second step this ``alternative'' relation is used to find the function $\phi(t)$ for $t<t_i$, and in the third step the relation (\ref{Eq:Lop_complete}) is finally found for $t<t_i$. Specifically, primarily the first two equations (\ref{Eq:radtransport}) are inserted into the relation (\ref{Eq:firstlaw}), obtaining 
\begin{align}\label{Eq:Lum_mass_der}
\frac{\partial L}{\partial m}=\,&\bar{\epsilon}-\frac{aT^4}{3}\>\frac{\partial }{\partial t}\left(\frac{1}{\rho}\right)-\frac{\partial }{\partial t}\left(\frac{aT^4}{\rho}\right)=\nonumber\\
=\,&\bar{\epsilon}+\frac{4aT^4}{3\rho}\> \left[\frac{\partial }{\partial t}\left(\log{\rho}\right)-3\frac{\partial }{\partial t}\left(\log{T}\right)\right]=\nonumber\\
=\,&\bar{\epsilon}-\frac{aT^4}{\rho}\> \left(\frac{d\log{\psi_t}}{d\omega}\frac{\partial \omega}{\partial t}+\frac{\dot{\phi}}{\phi}\right)=\nonumber\\
=\,&\bar{\epsilon}-\frac{aT^4}{\rho}\> \left[\left(1-\omega\>\cot{\omega}\right)\frac{\dot{x_i}}{x_i}+\frac{\dot{\phi}}{\phi}\right],
\end{align}
where $\dot{\phi}$ and $\dot{x}_i$ are the time derivatives of $\phi(t)$ and $x_i(t)$, respectively. Then, $L_{op}$ is inferred by integrating both members of relation (\ref{Eq:Lum_mass_der}) from the center of the ejected material up to the WCR mass coordinate $M_i(t)\equiv m[r(x_i,t)]= M_{ME}x_i^3(t)$ [cf.~relation (\ref{Eq:mass})], as follows:
\begin{align}
\label{Eq:Lopfirstlaw}
L_{op}(t)=\,&\int_0^{M_i(t)}\bar{\epsilon}dm+\nonumber\\
\,&-\frac{4a}{3\pi^2}R_i^3(t)\> \bigintsss_0^\pi T^4\times \left[\left(1-\omega\cot{\omega}\right)\frac{\dot{x_i}}{x_i}+\frac{\dot{\phi}}{\phi}\right]d\omega^3=\nonumber\\
=\,&\bigintsss_0^{M_i(t)}\bar{\epsilon}dm-\frac{E_{th}^0}{3\pi}\times\frac{x_i^3(t)\>\phi(t)}{(t/t_e+1)}\times\nonumber\\
\,& \times \bigintsss_0^\pi \left[\frac{\sin{\omega}}{\omega}\left(\frac{\dot{\phi}}{\phi}+\frac{\dot{x_i}}{x_i}\right)-\cos{\omega}\frac{\dot{x_i}}{x_i}\right]d\omega^3=\nonumber\\
=\,&\int_0^{M_i(t)}\bar{\epsilon}dm-E_{th}^0\>\frac{x_i^3(t)\phi(t)}{(t/t_e+1)}\> \left(\frac{\dot{\phi}}{\phi}+3\frac{\dot{x}_i}{x_i}\right).
\end{align}
At this point, the $L_{op}$ luminosity from relation (\ref{Eq:Lopfirstlaw}) can be compared with that from relation (\ref{Eq:Lop}) [or relation (\ref{Eq:Lop_dim_App})]. In particular, assuming $t>>t_e$ (or, similarly, $v_{sc}t>>R_0$), the comparison of the above mentioned relations gives the following equation:
\begin{equation}\label{Eq:FLMaster}
\frac{E_{th}^0}{t_d}\phi\> x_i\> \left[x_i^2\left(\frac{\dot{\phi}}{\phi}+3\frac{\dot{x}_i}{x_i}\right)+2\frac{t}{t_a^2}\right]=2\frac{t}{t_a^2}\> \int_0^{M_i(t)}\bar{\epsilon}dm.
\end{equation}
This equation can be used to derive the time evolution of $\phi$ during the diffusive phase. Indeed, using the relations $x_i(t)=1$ and $\dot{x}_i(t)=0$ that are valid in this phase [cf.~relation (\ref{Eq:boundcond1})], equation (\ref{Eq:FLMaster}) becomes
\begin{align}
\label{Eq:diffresults}
\dot{\phi}(t)=\,& -2\frac{t}{t_a^2}\> \left[\phi(t)-\frac{M_{ME}\> t_d}{E_{th}^0}\>\int_0^{1}\bar{\epsilon}(x,t)dx^3\right]=\nonumber\\
=\,& -2\frac{t}{t_a^2}\> \left[\phi(t)-\frac{S(t)}{E_{th}^0/t_d}\right]\quad \text{for } t<t_i \longrightarrow\nonumber\\
\phi(t)=\,&\frac{e^{-t^2/t_a^2}}{E_{th}^0/t_d}\>\left[\frac{E_{th}^0}{t_d}+\int_0^tS(t)d\left(e^{t'^2/t_a^2}\right)\right] \quad \text{for } t<t_i,
\end{align}
in which the latter is the general solution of $\phi$ for constant opacity models, where $S(t)$ can take into account any kind of internal heating sources [cf.~relation (\ref{Eq:source})]. Finally, inserting the relation (\ref{Eq:diffresults}) into the relation (\ref{Eq:Lop}) [or relation (\ref{Eq:Lop_dim_App})] and considering once again that $x_i(t<t_i)=1$, one obtains the relation
\begin{equation}
\label{Eq:Lop_complete_App}
L_{op}(t<t_i)=\left[ \frac{E_{th}^0}{t_d}+\bigintsss_{\>0}^t S(t')\,d\left(e^{t'^2/t_a^2}\right)\right]\>e^{-t^2/t_a^2},  
\end{equation}
which is equivalent to relation (\ref{Eq:Lop_complete}) for $t<t_i$ (Q.E.D.).\par

To derive relation (\ref{Eq:generalxi}), it is first necessary to equate relation (\ref{Eq:Lop}) with relation (\ref{Eq:boundcond2}). This translates to
\begin{align}
\label{Eq:reccond}
& E_{th}^0\> \phi(t) \> \frac{x_i(t)}{t_d}= 2\pi a cv_{sc}^2 T_{ion}^4 t^2\> x_i^2(t) \quad \text{for } t\geq t_i \quad\longrightarrow \nonumber\\
& \phi(t)=\frac{L_a\> t_d}{E_{th}^0}\> \left(\frac{t}{t_a}\right)^2\> x_i(t) \quad \text{for } t\geq t_i,
\end{align}
in which the latter links the time evolution of $\phi$ to the one of $x_i$ during the entire recombination phase. Then, substituting the relation (\ref{Eq:reccond}) into equation (\ref{Eq:FLMaster}),  considering relation (\ref{Eq:sourcei}), and supposing that $x_i$ is continuous at $t_i$ (see below for further details), one obtains
\begin{align}
\label{Eq:recresults}
& L_a\> t\> x_i^2\> \left[x_i^2\left(\frac{1}{t}+2\frac{\dot{x}_i}{x_i}\right)+\frac{t}{t_a^2}\right]= \int_0^{M_i(t)}\bar{\epsilon}dm \quad\text{with } x_i(t_i)=1 \quad\longrightarrow \nonumber\\
& 2t\>x_i^3\left(\Dot{x}_i+\frac{x_i}{2t}+\frac{t}{2t_a^2\>x_i}\right)=\frac{S_i(t)}{L_a}  \quad\text{with } x_i(t_i)=1  \quad\longrightarrow \nonumber\\
& x_i^2~\frac{d}{dt}\left[t\>\left(x_i^2+\frac{t^2}{3t_a^2}\right)\right]=\frac{S_i(t)}{L_a}  \quad\text{with } x_i(t_i)=1,
\end{align}
in which the latter is equivalent to relation (\ref{Eq:generalxi})  (Q.E.D.).\par

Last but not least, note that the couples of realtions (\ref{Eq:source})-(\ref{Eq:diffresults}) and (\ref{Eq:reccond})-(\ref{Eq:recresults}) define the temperature behavior of the ejected material during the first two post explosive phases (namely, diffusive and recombination phases). Since the transition between them at $t=t_i$ has to keep the temperature profile continuous, the relations
\begin{equation}\label{Eq:boundary_conditions}
\begin{cases}
x_i(t_i^-)=x_i(t_i^+)\\
\phi(t_i^-)=\phi(t_i^+)
\end{cases}
\longrightarrow
\begin{cases}
x_i(t_i^+)=1\\
\phi(t_i^-)=L_a\> \frac{t_d}{E_{th}^0}\> \left(\frac{t_i}{t_a}\right)^2
\end{cases}
\end{equation}
are valid, where the first relation on $x_i$ provides the initial boundary condition $x_i(t_i)=1$ for the Cauchy problem defined by the relation (\ref{Eq:generalxi}) and the second one on $\phi$ allows to find the beginning of the recombination according to the relation (\ref{Eq:ticond1}).

\section{Source functions and peculiar sub-models}\label{Appendix B}

In this appendix we present the \chem{56}{Ni} radioactive decay source term in more detail. In particular, introducing  the Ni-normalized source funtion $\bar{S}_i\equiv S_i/[M_{Ni}\>\epsilon(t)]$, all the sub-models defined in Section \ref{SSe:sourcef} can be distinguished thanks to it, as follows:
\begin{equation}
\label{Eq:HAR+SOE}
\bar{S}_i^{HAR+SOE}(t)=x_i^2(t)
\end{equation}
\begin{equation}
\label{Eq:HAR+IE}
\bar{S}_i^{HAR+IE}(t)=\frac{1-\sqrt{1-x_i^2}+x_i^2\left[1+log\left(\sqrt{1-x_i^{-2}}+x_i^{-1}\right)\right]}{2}
\end{equation}
\begin{equation}
\label{Eq:BOX+SOE}
\bar{S}_i^{BOX+SOE}(t)=
\begin{cases}
1\,& x_i(t)>x_c\\
\left[x_i(t)/x_c\right]^3\,& x_i(t)\le x_c
\end{cases}
\end{equation}
\begin{equation}
\label{Eq:BOX+IE}
\bar{S}_i^{BOX+IE}(t)=
    \frac{1}{2}\begin{cases}
        2\,& x_i(t)>x_c\\
        1+\left(\frac{x_i}{x_c}\right)^3 - \left[ 1-\left(\frac{x_i}{x_c}\right)^2\right]^{3/2} \,& x_i(t)\le x_c
    \end{cases}
\end{equation}
\begin{equation}
\label{Eq:EXP+SOE}
\bar{S}_i^{EXP+SOE}(t)=\frac{1-exp{[-k'_{mix}\>x_i^3(t)]}}{1-exp{[-k'_{mix}]}}.
\end{equation}
Only for the sub-model ``EXP+IE'' there is not an analytic Ni-normalized source function, essentially because, in this case, the term
$$\bigints x\> \sqrt{x^2-x_i^2}\> e^{-k'_{mix}x^3} \> dx$$
present in equation (\ref{Eq:IE}) is not expressible as a combination of elementary functions.\par

Once known the $\bar{S}_i$ (or, equivalently, the $S_i$) function, the $x_i$ time evolution can be found by solving numerically the Cauchy problem defined by the equation (\ref{Eq:generalxi}) for all the sub-models (cf.~Fig.~\ref{Fig:xiprofiles}). However, for the sub-model ``HAR+SOE'', there is also the following analytical solution: 
\begin{equation}
\label{Eq:solHAR+SOE}
x_i^2(t)=\frac{t_i}{t}\>\left(1+\frac{t_i^2}{3t_a^2}\right)-\frac{t^2}{3t_a^2}+q\> \frac{\Xi(t)-\Xi(t_i)}{t}
\end{equation}
with
\begin{equation}
\label{Eq:q_factor}
q=\frac{M_{Ni}\epsilon_{^{56}Co}\>\tau_{^{56}Co}}{L_a\>\left(\tau_{^{56}Co}-\tau_{^{56}Ni}\right)}\propto M_{Ni}\> M_{ME}^{-1/2}\> E^{-1/2}
\end{equation}
 and
\begin{align}
\Xi(t)=\,&\frac{\epsilon_{^{56}Ni}\>\tau_{^{56}Ni}}{\epsilon_{^{56}Co}\> \tau_{^{56}Co}}\>\left(\tau_{^{56}Co}-\tau_{^{56}Ni}\right)\times \left(1-e^{-t/\tau_{^{56}Ni}}\right) + \nonumber\\\
\,&+\left[\tau_{^{56}Co}\left(1-e^{-t/\tau_{^{56}Co}}\right)-\tau_{^{56}Ni}\left(1-e^{-t/\tau_{^{56}Ni}}\right)\right].
\end{align}

Moreover, since at the beginning of recombination most of \chem{56}{Ni} has already transformed into \chem{56}{Co} (indeed the relation $t_i\gtrsim \tau_{^{56}Ni}$ is typically verified), the relation (\ref{Eq:epsilon}) can be approximated as
\begin{equation}
\label{Eq:decay_approx}
\epsilon(t)\simeq \frac{\epsilon_{^{56}Co}\>\tau_{^{56}Co}}{\tau_{^{56}Co}-\tau_{^{56}Ni}}\times e^{-t/\tau_{^{56}Co}}= q\times \left(\frac{L_a}{M_{Ni}}\right)\times e^{-t/\tau_{^{56}Co}}
\end{equation}
and, consequently, for the sub-model ``EXP+SOE'' the differential equation in relation (\ref{Eq:generalxi}) can be rewritten in the following useful form:
\begin{equation}
\label{Eq:exp_diff}
\frac{dx_i}{dt}\simeq-\frac{x_i}{2t}-\frac{t}{2t_a^2x_i}+q'\> \left(\frac{1-e^{-k'_{mix}\>x_i^3}}{2t\>x_i^3}\right)\> e^{-t/\tau}  
\end{equation}
with $q'=q\times \left(1-e^{-k'_{mix}}\right)^{-1}$ and $\tau=\tau_{^{56}Co}$, hereafter.\par
Last but not least, note that in the case whether $M_{Ni}$ is equal to zero or the \chem{56}{Ni} heating effects on the WCR are neglected (i.e.~when the relation $S_i(t)= 0$ is verified), our model reproduces the one presented in \citet{popov93} as a special case. Indeed, for $S_i(t)= 0$, the equation (\ref{Eq:generalxi}) becomes homogeneous and can be rewritten as
\begin{equation*}
\frac{d}{dt}\left[t\>\left(x_i^2+\frac{t^2}{3t_a^2}\right)\right]=0.
\end{equation*}
So, by using the boundary condition $x_i(t_i)=1$, it can be analytically solved, obtaining the following relation:
\begin{equation}
\label{Eq:Popov_eq15}
t\>\left(x_i^2+\frac{t^2}{3t_a^2}\right)= t_i\>\left(1+\frac{t_i^2}{3t_a^2}\right), 
\end{equation}
which is equivalent to equation (15) of \citet{popov93} (Q.E.D.).

\section{Scaling Equations}\label{Appendix C}

In this appendix, we focus on the derivation of three purely photometric scaling equations that relate the peak features of 1987A-like SNe to parameters describing the SN progenitor at explosion. To this aim, it is useful to rewrite equation (\ref{Eq:L_SN}) in an appropriate manner. In particular, during the peak phase (i.e.~$30\,\text{d}\lesssim t\lesssim t_f$) $L_{SN}$ is dominated by the ME luminosity, therefore equation (\ref{Eq:L_SN}) becomes
\begin{equation}
\label{Eq:lumSN_peak}
L_{SN}(t)\simeq L_{op}+L_{tr}=L_a\left(\frac{t}{t_a}\right)^2\,x_i^2(t)+ M_{Ni}\>\epsilon(t) \left[1-\bar{S}_i(t)\right],
\end{equation}
where equation (\ref{Eq:ME_luminosity}) has been substituted for $L_{ME}$, and $L_{op}$ and $L_{tr}$ have been respectively expressed through relations (\ref{Eq:boundcond2}) and (\ref{Eq:lumtra}), using the quantities $L_a$ and $\bar{S}_i$ defined in note \ref{note_ti} and Appendix \ref{Appendix B}. The equation (\ref{Eq:lumSN_peak}) is thus a general expression for $L_{SN}$ from which the scaling relations during the peak phase can be derived.\par

Starting with the rising phase (i.e.~$30\,\text{d}\lesssim t\lesssim \tilde{t}$), the BOX approach allow us to simplify the equation (\ref{Eq:lumSN_peak}) replacing the expressions (\ref{Eq:BOX+SOE})-(\ref{Eq:BOX+IE}) to $\bar{S}_i$, so
\begin{equation}
\label{Eq:Lum_BOX}
L_{SN}^{BOX}(t)=L_a\left(\frac{t}{t_a}\right)^2\,x_i^2(t)
\end{equation}
as long as $x_i(t)\ge x_c$, which is valid for $t \lesssim \tilde{t}$ [cf.~Figs \ref{Fig:xiprofiles} and \ref{Fig:LC_features}]. Furthermore, by comparing the profiles of $x_i$ in Fig.~\ref{Fig:xiprofiles}, we observe that the behaviour of the $x_i$-curves for all our sub-models are similar within the first 40-50 days at least and, in any case, for $t \lesssim \tilde{t}$. Therefore, in equation (\ref{Eq:Lum_BOX}) we can substitute the analytical form of $x_i$ found for the HAR case and, neglecting in equation (\ref{Eq:solHAR+SOE}) the terms that depend on $t_a^{-2}$ (because the relation $t_i<\tilde{t}<<t_a$ is generally valid for 1987A-like SNe), we obtain:
\begin{equation}
\label{Eq:Lum_BOX_xiHar}
L_{SN}^{BOX}(t)\simeq L_a\times\frac{t\,t_i}{t_a^2}\times\left(1+q\> \frac{\Xi(t)-\Xi(t_i)}{t_i}\right).
\end{equation}
Since here $t$ and $t_i$ are generally greater than $\tau_{^{56}Ni}$, we can also use the approximation of equation (\ref{Eq:decay_approx}), thus the difference between the values of $\Xi$ calculated in $t$ and $t_i$ can be simplified as follows
\begin{align}
\Xi(t)-\Xi(t_i)=\,&\frac{M_{Ni}}{q\,L_a}\times\int_{ t_i}^{t}\epsilon(t')\,dt'\nonumber\\
\simeq\,&\tau_{^{56}Co}\times\left[e^{-t_i/\tau_{^{56}Co}}-e^{-t/\tau_{^{56}Co}}\right]\nonumber\\
\simeq\,& t-t_i+o(t/\tau_{^{56}Co}),
\end{align}
in which the exponentials have been approximated to the first order since $t_i<t\le\tilde{t}$ are one order of magnitude smaller than $\tau_{^{56}Co}$. In light of this, the derivative of the magnitude during the rising phase can be expressed as follows:
\begin{align}\label{Eq:risingM}
 \dot{M}_{rising}(t)\equiv\,&\left.\frac{dM}{dt}\right|_{t\in[30\,\,\text{d},\tilde{t}]}=-2.5\times \frac{d\log_{10}L_{SN}^{BOX}}{dt}\nonumber\\
 \simeq\,&-(2.5/\ln{10})\times\left[\frac{1}{t}+\frac{q}{t_i}\right]+ o\left(q\,\frac{t-t_i}{t_i}\right).
\end{align}
By using the relation (\ref{Eq:risingM}), we can express $D\tilde{M}$ defined in equation (\ref{Eq:DM_tilde}) to find the first scaling relationship on the rising phase for 1987A-like SNe:
\begin{equation}\label{Eq:scaling1_app}
D\tilde{M}\simeq -(2.5/\ln{10})\times \frac{q}{t_i}\propto M_{Ni}\times (E\,M_{ME}\,R_0)^{-1/2}\quad\text{(Q.E.D.).}
\end{equation}

In order to facilitate the analysis of the LC behaviour for the subsequent phases and find other scaling relations, it is useful to introduce $\bar{L}$ as the ratio between the SN luminosity and \chem{56}{Ni} tail ones (i.e.~$\bar{L}\equiv L_{SN}(t)/[M_{Ni}\,\epsilon(t)]$). According to the approximation of relation (\ref{Eq:decay_approx}), $\bar{L}$ can be written as follows: 
\begin{align}
\label{Eq:L_bar}
\bar{L}(t)= 1+\frac{L_a\,(t/t_a)^2}{M_{Ni}\,\epsilon(t)}\,x_i^2(t)-\bar{S}_i(t)\simeq 1+b(t)\,x_i^2(t)-\bar{S}_i(t),
\end{align}
where $b(t)=q^{-1}\times \left(t/t_a\right)^2\,e^{t/\tau}$ is the ratio of the BB brightness with radius $R(t)=v_{sc}\> t$ and surface temperature $T_{ion}$ to the radioactive luminosity of a \chem{56}{Ni} transparent sphere with mass $M_{Ni}$. With the $\bar{L}$ introduction, the difference between the SN magnitude (i.e.~$M_{SN}=-2.5\, \log_{10}L_{SN}+88.7$) and $M_{tail}$ [the same of equation (\ref{Eq:M_tail}) with $m_{Ni}=-2.5\, \log_{10}M_{Ni}+88.7$] can be expressed as
\begin{equation}
\label{Eq:Delta_M}
\Delta M\equiv M_{SN}-M_{tail}=-2.5\,\log_{10}\bar{L}.
\end{equation}
This theoretical curve is dependent upon the sub-model type and modeling parameters. Moreover, $\Delta M$ can be used to simulate the features of $M_{bol}-M_{tail}$ obtained by the LC data analysis (cf. Section \ref{SubSec:quarta_3}). \par
Analogously, we can define the time-derivative of $\Delta M$, which becomes
\begin{equation}\label{Eq:D_M}
DM\equiv \frac{d(\Delta M)}{dt}=\dot{M}_{SN}-\dot{M}_{tail}=-(2.5/\ln{10})\times \bar{L}^{-1}\times \frac{d\bar{L}}{dt}.
\end{equation}
In the SOE case, $d\bar{L}/dt$ can be written by deriving respect to $t$ the equation (\ref{Eq:L_bar}) as follows
\begin{align}\label{Eq:L_bar_deriv}
\frac{d\bar{L}}{dt}\simeq\,& b(t)\,x_i^2\,\left(\frac{2}{t}+\frac{1}{\tau}+2\,\frac{\dot{x}_i}{x_i}\right)-3\,\xi(x_i)\,x_i^2\,\dot{x}_i=\nonumber\\
=\,&t^{-1}\,\left\{\left[\left(1+\frac{t}{\tau}\right)\,b(t)+\frac{3}{2}\,\xi(x_i)\,x_i\right]\,x_i^2+\right.\nonumber\\
\,&\left.-q\,e^{-t/\tau}\,\left[b(t)-\frac{3}{2}\,\xi(x_i)\,x_i\right]\,\left[b(t)-\bar{S}_i(t)\,x_i^{-2}\right]\right\},
\end{align} 
where the derivative of $\bar{S}_i$ has been substituted for $3\, \xi[x_i(t)]\, x_i(t)^2\> \dot{x}_i$ [cf.~equation (\ref{Eq:SOE})], thus making equation (\ref{Eq:L_bar_deriv}) valid for all sub-models under the SOE condition. Furthermore, in equation (\ref{Eq:L_bar_deriv}) the dependence on the derivative of $x_i$ has been removed by getting $\dot{x}_i$ from equation (\ref{Eq:generalxi}) [cf.~equation (\ref{Eq:recresults})].\par

In order to link the features of the LC secondary maximum, such as $\bar{t}_M$ and $\bar{L}_M$ (cf. Section \ref{SubSec:quarta_3}), to the modelling parameters in a direct way, it is necessary to use the sub-model ``HAR+SOE''. Indeed, in this case both equations (\ref{Eq:L_bar}) and (\ref{Eq:L_bar_deriv}) can be simplified as follows:
\begin{equation}
\label{Eq:lum_ratioHAR}
\bar{L}^{HAR}(t)=1+\left[b(t)-1\right]\> x_i^2,
\end{equation}
\begin{align}\label{Eq:lum_derHAR}
\frac{d\bar{L}^{HAR}}{dt}=\,&\left[\left(t^{-1}+\tau^{-1}\right)\,b(t)
+(3/2)\,t^{-1}\right]\,x_i^2\nonumber\\
&-\frac{q\,e^{-t/\tau}}{t}\,\left[b(t)-3/2\right]\,\left[b(t)-1\right].
\end{align}
In this way, $\bar{t}_M$ and $\bar{L}_M$ can be related by the following analytic expression:
\begin{equation}\label{Eq:Lbar_MHAR}
\bar{L}_M\simeq \bar{L}^{HAR}(\bar{t}_M)=1+\left[b(\bar{t}_M)-1\right]\> x_i^2(\bar{t}_M),
\end{equation}
where 
\begin{equation}\label{Eq:xi_MHAR}
x_i^2(\bar{t}_M)=q\,e^{-\bar{t}_M/\tau}\times\frac{\left[b(\bar{t}_M)-3/2\right]\,\left[b(\bar{t}_M)-1\right]}{\left[\left(1+\bar{t}_M/\tau\right)\,b(\bar{t}_M)
+3/2\right]}
\end{equation}
is derived from the condition $DM(\bar{t}_M)=0\rightarrow d\bar{L}^{HAR}/dt|_{\bar{t}_M}=0$ [in accordance with equation (\ref{Eq:D_M}), cf.~also note \ref{Note:t_barM}]. Furthermore, at $\bar{t}_M$ the BB brightness given by $L_a\,(\bar{t}_M/t_a)^2$ is significantly higher than the Ni luminosity $M_{Ni}\epsilon(\bar{t}_M)$, therefore their ratio expressed by $b(\bar{t}_M)$ has to be grater than one [e.g.~$b(\bar{t}_M)\simeq19$ for SN 1987A]. By approximating the equations (\ref{Eq:Lbar_MHAR})-(\ref{Eq:xi_MHAR}) for $b(\bar{t}_M)>>1$, we can simplify as follows:
\begin{align}
\bar{L}_M\simeq\,& 1+q\,e^{-\bar{t}_M/\tau}\times\frac{\left[b(\bar{t}_M)-3/2\right]\,\left[b(\bar{t}_M)-1\right]^2}{\left[\left(1+\bar{t}_M/\tau\right)\,b(\bar{t}_M)
+3/2\right]}\nonumber\\
\simeq\,&1+\frac{q\,e^{-\bar{t}_M/\tau}}{1+\bar{t}_M/\tau}\times \left[b(\bar{t}_M)\right]^2\simeq 1+ (q\,t_a^4)^{-1}\,\frac{\bar{t}_M^4\,e^{\bar{t}_M/\tau}}{1+\bar{t}_M/\tau}\nonumber\\
\simeq\,&1+(q\,t_a^4)^{-1}\,\bar{t}_M^4+o(\bar{t}_M/\tau),
\end{align}
from which, by neglecting the higher-order terms in $\bar{t}_M/\tau$, we can obtain the second scaling equation
\begin{equation}\label{Eq:scaling2_app}
 \bar{t}_M^4\,(\bar{L}_M-1)^{-1}\simeq (q\,t_a^4)\propto M_{Ni}\times (E^{-3}\,M_{ME}^5)^{1/2}\quad(Q.E.D.).
\end{equation}

The last scaling relation concerns the final stage of recombination, which occurs at $t_f$ in our model. At this epoch, $DM$ presents a discontinuity of the first type that makes it jump from the value
\begin{equation}\label{Eq:DM_f}
DM_f=-(2.5/\ln{10})\,\left.\frac{d\bar{L}}{dt}\right|_{t_f^-}
\end{equation}
to zero, because the relation $M_{SN}=M_{tail}$ is valid from the recombination end time onwards. The left derivative of $\bar{L}$ (i.e.~$d\bar{L}/dt|_{t_f^{-}}$) can be calculated rigorously via the equation (\ref{Eq:L_bar_deriv}). Indeed, assuming a \chem{56}{Ni} density profile such that $\xi(x)\sim x^{p}$ around to $x=0$ with $p>-1$ (note that in this way only the HAR case is not included), the continuity condition of $x_i(t)$ gives:
\begin{equation}
\lim_{t\rightarrow t_f^{-}}x_i(t)=\lim_{t\rightarrow t_f^{-}}\xi[x_i(t)]\,x_i(t)=\lim_{t\rightarrow t_f^{-}}\bar{S}_i(t)\,x_i^{-2}(t)=0,
\end{equation}
so, calculating the same limit for the equation (\ref{Eq:L_bar_deriv}), we have
\begin{equation}\label{Eq:scaling3_prev}
\left.\frac{d\bar{L}}{dt}\right|_{t_f^-}\equiv \lim_{t\rightarrow t_f^{-}}\frac{d\bar{L}}{dt}\simeq\,e^{-t_f/\tau}\,[b(t_f)]^2=-\frac{t_f^3\> e^{t_f/\tau}}{qt_a^4},
\end{equation}
from which we obtain the last scaling equation
\begin{align}\label{Eq:scaling3_app}
DM_f\propto\,& \frac{t_f^3\> e^{t_f/\tau}}{M_{Ni}\> (M_{ME}^{5}\> E^{-3})^{1/2}}\longrightarrow\nonumber\\
\,& DM_f^{-1}\> t_f^3\> e^{t_f/\tau}\propto M_{Ni}\times (E^{-3}\,M_{ME}^5)^{1/2} \quad (Q.E.D.).
\end{align}

\label{lastpage}
\end{document}